\documentclass[traditabstract]{template/aa} 

\usepackage{template/natbib}
\usepackage{array}
\usepackage{float}
\usepackage[fleqn]{amsmath}
\usepackage{amssymb}
\usepackage{stfloats}
\usepackage{morefloats}
\usepackage{multirow}
\bibpunct{(}{)}{;}{a}{}{,}
\usepackage[dvips]{graphicx}
\usepackage{slashbox}
\usepackage{xfrac}
\usepackage{hyperref}
\usepackage{breakurl}
\usepackage{txfonts}
\usepackage{breqn}

\usepackage{pstricks}
\usepackage{caption}
\usepackage{subcaption}
\usepackage{color}
\usepackage{breqn}
\usepackage[shortcuts]{extdash}
\begin{document}
   \title{Dust grains from the heart of supernovae}

   \author{M. Bocchio\inst{1, 2}, 
          S. Marassi\inst{2},
          R. Schneider\inst{2},
          S. Bianchi\inst{1},
          M. Limongi\inst{2},
          A. Chieffi\inst{3}}
          
   \institute{INAF - Osservatorio Astrofisico di Arcetri, Largo Enrico Fermi 5, 50125 Firenze, Italy\\
   	INAF - Osservatorio Astronomico di Roma, Via di Frascati 33, I-00040 Monteporzio, Italy\\
	INAF/IASF, Via Fosso del Cavaliere 100, 00133 Roma, Italy\\                          
				}
 
  \abstract
  {Dust grains are classically thought to form in the winds of Asymptotic Giant Branch (AGB) stars.
  However, nowadays there is increasing evidence for dust formation in Supernovae (SNe). In order
  to establish the relative importance of these two classes of stellar sources of dust it is important
  to know what is the fraction of freshly formed dust in SN ejecta that is able to survive the
  passage of the reverse shock and be injected in the interstellar medium.
  
   With this aim, we have developed a new code, GRASH\_Rev, that allows to 
   follow the dynamics of dust grains in the shocked SN ejecta and to compute the time
   evolution of the mass, composition and size distribution of the grains. We consider 
   four well studied SNe in the Milky Way and Large Magellanic Cloud: SN 1987a, Cas A,
   the Crab Nebula, and N49. These sources have been observed with both {\it Spitzer}
   and {\it Herschel} and the multiwavelength data allow to better assess the mass of warm
   and cold dust associated with the ejecta. 
 
For each SN, we first identify the best explosion model, using the mass and metallicity 
of the progenitor star, the mass of $^{56}$Ni, the explosion energy and
the circumstellar medium density inferred from the data. We then run a dust formation model to 
compute the properties of freshly formed dust (Marassi et al. 2015).
Starting from these input models, GRASH\_Rev self-consistely follow the dynamics of the grains
considering the effects of the forward and reverse shock and allows to predict the time evolution
of the dust mass, composition and size distribution in the shocked and unshocked regions of the ejecta.
  
For all the simulated models, we find good agreement with observations.  
Our study suggests that SN 1987A is too young for the reverse shock to have affected the dust mass. Hence
the observed dust mass of $0.7 -0.9\, M_\odot$ in this source can be safely considered as indicative
of the mass of freshly formed dust in SN ejecta. Conversely, in the other three SNe, the reverse
shock has already destroyed between 10 and 40\% of the initial dust mass. 
However, the largest dust mass destruction is predicted to occur between $10^3$ and $10^5$\, yr after
the explosions. Since the oldest SN in the sample has an estimated age of 4800 yr, 
current observations can only provide an upper limit to the mass of SN dust that will enrich
the interstellar medium, the so-called effective dust yields. 
We find that only between 1 and 8\% of the currently observed mass will survive, resulting in an  average SN effective
dust yield of $(1.55 \pm 1.48)\times 10^{-2} \, M_\odot$. This is in good agreement with the values
adopted in chemical evolution models which consider the effect of the SN reverse shock. 

We discuss
the astrophysical implications of our results for dust enrichment in local galaxies and at high redshift.}


   \maketitle
%

\section{Introduction}

It is observationally and theoretically well established that a considerable amount of
dust is efficiently formed in regions around Asymptotic Giant Branch (AGB) stars 
(e.g. \citealt{2002A&A...392..131W,2006A&A...452..537W,2008AA...479..453Z,2010A&A...515A..27O,2014MNRAS.442.1440S,2015MNRAS.447.2992D}).
This process is classically considered as the primary source of dust grains 
in galaxies (see however Schneider et al. 2014) and in the Milky Way the typical formation timescale is $\sim 3\times 10^9$\,yr 
(\citealt{1980ApJ...239..193D,1989IAUS..135..445G,1994dib..nasa...79J}).

On the contrary, supernova (SN) explosions in the interstellar medium (ISM) trigger 
shock waves which are able to quickly process dust grains and are considered
the dominant mechanism of dust destruction in the ISM 
(\citealt{1978MNRAS.183..397B,1978MNRAS.183..367B,1979ApJ...231..438D,1979ApJ...231...77D,1980ApJ...239..193D,1983ApJ...275..652S,1987ApJ...318..674M,1994ApJ...433..797J,1996ApJ...469..740J, 2015ApJ...799...50L, 2015ApJ...803....7S}).
A recent theoretical work on interstellar dust destruction in shock waves led to an
estimated lifetime of $\sim 6 \times 10^7$\,yr and $\sim 3\times 10^8$\,yr for carbonaceous and
silicate grains in our Galaxy, respectively (Bocchio et al. 2014), which is much shorter
than the assumed dust formation timescale from AGB stars.
This leads to the conclusion that a large amount of dust must be re-accreted from the
gas phase.
However, while there seems to be viable mechanisms for re-formation of carbonaceous grains
under low temperature and pressure conditions (e.g. \citealt{2005A&A...432..895D}),
silicate grains are harder to form under ISM conditions (\citealt{2011A&A...530A..44J}).

\begin{table*}[t!]
\caption{Observed/estimated physical properties of the four SNe considered in the present study
that we have compiled from the papers listed in column ``Ref''.
The last three columns indicate the input parameters used in the simulations.\vspace{.2cm}}
\label{table:par_sample}      
\centering          
\begin{tabular}{c | c c c c c c c c c | c c c }
\hline\hline
SNR & type &  $E_{\rm ex} $& $^{56}$Ni & $M_{\rm prog}$ & Z& Age & $n_{\rm 0}$& $M_{\rm dust}$ & Ref. & $E_{\rm ex}^{\rm mod} $& $^{56}$Ni$^{\rm mod}$ & $M_{\rm prog}^{\rm mod}$ \\
 & & [$10^{51}$ erg] & [$M_{\odot}$] & [$M_{\odot}$] & [$Z_{\odot}$] & [Years] & [cm$^{-3}$] & [$M_{\odot}$] & & [$10^{51}$ erg] & [$M_{\odot}$] & [$M_{\odot}$]\\
\hline
1987A & II$^{\rm a}$ & 1.1 & 0.075$^{\rm a}$ & 18-20$^{\rm a}$ & 0.4 & 27 & 1.5 & 0.7-0.9 & g, h, i  &1.0 & 0.075 & 20\\
Cas A & IIb$^{\rm b}$ & 1-3 & 0.058-0.16$^{\rm c}$ & 13-20$^{\rm b}$ & 1 & 335 & 1.9 & 0.1-1& j, k, l  &1.5 & 0.11 & 20\\
Crab & IIn-P$^{\rm d}$ & 0.1 & 0.009 - 0.016$^{\rm d}$ & 9-12$^{\rm e}$& 1 & 960 & 1 & 0.1-0.25& m &0.5 & 0.014 & 13\\
N49 & II & 1.8 & N/A & 18-20 & 0.4 & 4800 & 0.9 & 0.1-0.4& n, o &1.5 & 0.075 & 20\\
\hline
\end{tabular}
\\\vspace{2mm} 
References: $^{\rm a}$\cite{1989ARA&A..27..629A},$^{\rm b}$\cite{2008Sci...320.1195K}, 
$^{\rm c}$\cite{2009ApJ...697...29E},$^{\rm d}$\cite{2013MNRAS.434..102S}, $^{\rm e}$\cite{2008AJ....136.2152M},
$^{\rm f}$\cite{2014MNRAS.441.1040R}, $^{\rm g}$\cite{2011Sci...333.1258M}, $^{\rm h}$\cite{2014ApJ...782L...2I}, $^{\rm i}$\cite{2011A&A...532L...8L}, $^{\rm j}$\cite{2010A&A...518L.138B},
$^{\rm k}$\cite{2008ApJ...673..271R}, $^{\rm l}$\cite{2009MNRAS.394.1307D}, $^{\rm m}$\cite{2012ApJ...760...96G}, $^{\rm n}$\cite{2012ApJ...748..117P}, $^{\rm o}$\cite{2010A&A...518L.139O}.
\end{table*}

Although SNe are believed to be efficient interstellar dust destroyers, nowadays there is increasing 
observational evidence for the formation of non-negligible quantities of dust grains
in the ejecta of SNe (e.g. \citealt{2009MNRAS.396..918M,2009MNRAS.394.1307D,2011A&ARv..19...43G,2014Natur.511..296G}).
Given the relatively short timescale between the explosion of two SNe (estimated to be $\sim 40$\,yr in our 
Galaxy, \citealt{2011MNRAS.412.1473L}), this would lead to an effectively shorter timescale 
for dust formation. 

Rapid dust enrichment is also required by millimeter (mm) and sub-millimeter (submm) observations
of $z > 6$ galaxies \citep{2015Natur.519..327W} and quasars (see \citealt{2014MNRAS.444.2442V} and references therein), 
which show that their ISM has already been enriched with $\gtrsim 10^8 M_\odot$ of dust.  Theoretical models
and numerical simulations show that these large dust masses require efficient grain growth in the dense
phases of the ISM (\citealt{2014MNRAS.444.2442V,2015A&A...577A..80M,2015MNRAS.451L..70M}), with both SN and AGB 
stars providing the first seed grains \citep{2011MNRAS.416.1916V}.

One important source of uncertainty on the role of SNe as efficient dust polluters is whether the dust formed in the ejecta survives
 the passage of the reverse shock. In fact, the observed supernova remnants (SNR) are still relatively young (age $\lesssim 5000$\,yr) and
the reverse shock has not reached the centre of the ejecta yet. The measured mass of 
dust associated to the ejecta will therefore not be entirely released into the ISM
but partly destroyed.

Only in four core-collapse SNe it has been possible
to estimate the amount of dust in the ejecta, using observation from
Spitzer and Herschel: SN 1987A, Cassiopeia A, 
Crab Nebula and SNR N49.
The photometer instruments on these two satellites, covering
together a large range of wavelengths from Mid-Infrared (MIR) through 
Far-Infrared (FIR) and submm,
are able to measure the emission from both cold and warm dust in SNe.
However, in order to estimate the dust mass responsible for the detected emission, correct assumptions
must be made on the dust composition, size and temperature distribution.
Depending on the choice of the dust composition and on the size/temperature of the smallest grains,
the dust mass estimate can vary by almost a factor of 6 (\citealt{2015MNRAS.449.4079M}).
Furthermore, emission from cold dust in the molecular cloud surrounding a SN contaminates the
emission from freshly-formed dust in the ejecta and different observations (e.g. polarisation measurements) 
are therefore needed to disentangle the two components \citep{2009MNRAS.394.1307D}.
The difficulty in determining how much emission comes from dust in the
remnant and how much from the ISM was also recognised in the recent analysis 
of a sample of SNRs in the LMC by \cite{2015ApJ...799...50L}: while they put constraints 
on the influence of the remnant on the nearby ISM, the data lacked the resolution and 
sensitivity to study the amount of dust within the remnants themselves.

In order to estimate the dust mass which is released into the ISM
after the passage of the reverse shock, a numerical code that simulates the dust dynamics and 
processing in SNe is needed.
Using numerical simulations, \cite{2006ApJ...648..435N,2007ApJ...666..955N}  estimated that, depending on the energy of the explosion
and on the density of the ISM, the fraction of dust surviving the reverse shock can vary between 0 and 80~\%.
\citet[BS07]{2007MNRAS.378..973B} estimated that depending on
the density of the surrounding interstellar medium, between 2 and 20~\% of the initial
dust mass survives the passage of the reverse shock.

In this work we present a new code called GRASH\_Rev that treats dust processing in a supernova 
explosion. This code couples all the dust processing included in the GRASH\_EX code (\citealt{2014A&A...570A..32B})
with the dynamics and structure of the SN as modelled by BS07 but extending it 
to include the full dynamics of dust grains within the ejecta and in the surrounding ISM.

This paper is organised as follows: in Section~\ref{sect:sample} we first present the SN sample that we have
considered to compare the model predictions with observational data. In Section~\ref{sect:structure_ejecta} we describe
the corresponding SN explosion models and the properties of dust formed in their ejecta.
In Section~\ref{sect:GRASH_Rev} we present the structure of the numerical model, the grid used in
the calculation, how we describe gas and dust dynamics, and the physical processes
included in dust processing. In Sections~\ref{sect:results} and \ref{sect:discussion} we present the main results and we
discuss their astrophysical implications. Finally, in 
Section~\ref{sect:conclusions} we draw our conclusions.

\section{Supernova sample}
\label{sect:sample}

We consider the four aforementioned core-collapse SNRs:
SN 1987A, Cassiopeia A, Crab nebula and SNR N49.
In Table~\ref{table:par_sample} we report some of their physical properties, as obtained from
the literature:
the type of SN, the explosion energy ($E_{\rm ex}$), the mass of $\rm ^{56}Ni$, the progenitor mass ($M_{\rm prog}$),
the estimated metallicity ($Z$) of the progenitor star, the age, the number density of the circumstellar ISM ($n_{\rm 0}$)
and the measured dust mass associated with the ejecta ($M_{\rm dust}$).
Some of these quantities are directly estimated from the data, while others are obtained by
comparing the observed properties with theoretical models. For example, 
the progenitor mass and explosion energy of SN 1987A are obtained by fitting the observed light curve and spectrum 
with a theoretical model (\citealt{1994supe.conf..489N, 2006NuPhA.777..424N}). 
Similarly, the mass of $^{56}$Ni is estimated from comparison between theoretical and observed light curves. 
The uncertainties / approximations of the models are encoded in the final estimates of these physical quantities.

\subsection{SN 1987A}

SN 1987A is a young SN located in the Large Magellanic Cloud (LMC) and exploded in 1987.
Observational data in Table~\ref{table:par_sample} are from \cite{2013lcdu.confE.146G} and 
references therein.
The dust mass estimate has been updated following the recent work by \cite{2015ApJ...800...50M}, leading to a total 
dust mass of $\sim 0.8\,{\rm M}_{\odot}$.
However, a recent and new analysis of the infrared emission from SN1987A by \cite{2015arXiv150707019D} points to
a lower dust mass of $\sim 0.45\,{\rm M}_{\odot}$, mostly composed by silicate grains.

\subsection{Cassiopeia A}

\begin{table*}[t!]
\caption{Dust distribution parameters and masses for the four SNe in the sample.\vspace{.2cm}}
\label{table:par_sample2}      
\centering          
\begin{tabular}{@{\extracolsep{1mm}} l c c c c  c c c c  c c c c }
\hline\hline
  &  \multicolumn{3}{c}{1987A}   & \multicolumn{3}{c}{CasA}  & \multicolumn{3}{c}{Crab}& \multicolumn{3}{c}{N49}\\
\cline{2-4}
\cline{5-7}
\cline{8-10}
\cline{11-13}
  & $a_0$ & $\sigma$ & $M_{\rm dust}$   & $a_0$ & $\sigma$ & $M_{\rm dust}$   & $a_0$ & $\sigma$ & $M_{\rm dust}$   & $a_0$ & $\sigma$ & $M_{\rm dust}$ \\
  & (nm) & & ($M_{\odot}$) & (nm) & & ($M_{\odot}$) & (nm) & & ($M_{\odot}$) & (nm) & & ($M_{\odot}$) \\
\hline
${\rm Al_2O_3}$ &7.5 & 0.14 & $2.57\times10^{-2}$  &5.5 & 0.14 & $2.57\times10^{-2}$ & 4.7 & 0.01 & $5.18\times10^{-3}$&5.5 & 0.14 & $2.57\times10^{-2}$\\
${\rm Fe_3O_4}$ &9.3 & 0.13& 0.11 &9.5 & 0.13 & 0.15 &6.1 & 0.01&$2.85\times10^{-2}$&7.0 & 0.13& 0.11 \\
${\rm MgSiO_3}$ &84.9 & 0.08&$6.58\times10^{-6}$&61.6 & 0.08 & $2.67\times10^{-5}$&3.0 & 0.11&$2.68\times10^{-3}$&61.5 & 0.08& $2.74\times10^{-5}$\\
${\rm Mg_2SiO_4}$ &68.9 & 0.11& 0.43  &50.5 & 0.11& 0.43&4.3 & 0.13& $8.60\times10^{-2}$&50.4 & 0.11& 0.43\\
${\rm AC}$ &90.4 & 0.13&$7.15\times10^{-2}$ &105.2 & 0.13& 0.12&26.0 & 0.14&0.11&103.2 & 0.13& 0.12\\
${\rm SiO_2}$ &55.5 & 0.16&0.19 &38.9 & 0.16& 0.187&4.5 & 0.03 & $1.11\times10^{-2}$&39.0 & 0.16& 0.187\\
\hline
$M_{\rm TOT}$  &&&0.84 && & 0.92& &&0.245&&& 0.88\\
\hline
\end{tabular}
\end{table*}

Cassiopeia A (CasA) is a Galactic SNR resulting from a SN explosion occurred 335 years ago.
The kinetic energy has been estimated to be $\sim (2 - 3) \times 10^{51}$erg (\citealt{2003ApJ...597..347L}).
The mass of the shocked and unshocked ejecta is $\sim 3-4$ M$_{\odot}$ (\citealt{2012ApJ...746..130H}) and
the mass of gas swept away in the circumstellar medium is $\sim 8$ M$_{\odot}$ (\citealt{2009ApJ...703..883H,2009ApJ...697..535P}).
Adding the mass of a neutron star remnant, the total mass of the progenitor is estimated to be $\sim (15 \pm 3)$\,M$_{\odot}$.
The mass of dust associated to the SNR inferred both from {\it Herschel} observations
and 850 $\mu$m polarimetry data  is $M_{\rm dust} = 0.1 - 1$\,M$_{\odot}$.
Given the large uncertainty in the mass of $^{56}$Ni, we adopt an intermediate value as input parameter of the simulation.

\subsection{Crab nebula}

The small progenitor mass (8-10 $M_{\odot}$), the low explosion energy 
($10^{50}$ erg) and mass of synthesized $^{56}$Ni (0.009-0.016 M$_{\odot}$), which are 
significantly less than the energy and nickel mass expected for ordinary SNe, all
suggest that the Crab could be the result of an electron-capture SN event (\citealt{2013MNRAS.434..102S}). 
The adopted  $^{56}$Ni mass for the simulation is intermediate between the observed ones.

\subsection{SNR N49}
The estimated physical properties of N49 have been taken from \cite{2012ApJ...748..117P}.
\citet{2010A&A...518L.139O} reports a mass of warm dust of $\sim 0.1 - 0.4$\,M$_{\odot}$. 
Emission from a large mass of cold dust is also observed, but it is not possible to disentangle the contribution coming from 
dust in the unshocked ejecta from that in the parent cloud.

For this source, there is no estimate of the $^{56}$Ni mass available. We therefore 
adopt the same value used to model SN 1987A.

\section{Structure of the ejecta and grain distribution}
\label{sect:structure_ejecta}

Starting from a homogeneous set of solar metallicity presupernova models with masses in the range [13-130]~$M_\odot$ 
\citep{2013ApJ...764...21C} simulated by means of the FRANEC stellar evolutionary code 
(\citealt{2006ApJ...647..483L}), we have selected the most suitable model for each SN according to the physical properties listed in Table 1. 
These four selected models are then used as input for the dust formation code (BS07), 
where classical nucleation theory in steady state conditions was applied. Here we use the latest version of dust formation model, which 
implements an upgraded molecular network, as described in detail in \citet{Marassi2015}. We follow the formation
of six different dust grain species: amorphous carbon (AC), corundum (Al$_2$O$_3$), magnetite (Fe$_3$O$_4$), enstatite (MgSiO$_3$), forsterite (Mg$_2$SiO$_4$) 
and quartz (SiO$_2$). 

In Table~\ref{table:par_sample2} we show the resulting dust composition and mass distribution.
The grains that condense and accrete in SN ejecta follow a log-normal size distribution function
of the form,
\begin{equation}
m_i(a) = m_{{\rm max}, i} \exp \left[- \frac{\log_{10}^2(a/a_{0, i})}{2{\sigma_i}^2}\right],
\end{equation}
\noindent
where $m_{{\rm max}, i}$ represents the distribution maximum, $a_{0, i}$ is the centroid, ${\sigma_i}$ is the width
of the distribution and
\begin{equation}
M_{{\rm dust}, i} = \int_0^{\infty} m_{i}(a)\,da,
\end{equation}
is the total dust mass for the $i$-th grain species.
Table~\ref{table:par_sample2} shows $a_0$, $\sigma$ and $M_{\rm dust}$
for the six dominant grain species in each SN.

In this work we consider a carbon grain density $\rho = 2.2$\,g cm$^{-3}$. However, presolar grain measurements
indicate a wide range of density for carbonaceous grains formed in SNe (e.g. \citealt{2005LPI....36.1867A,2014AIPC.1594..307A}).
We explore this scenario in Appendix~\ref{app:carb_dens}.

\section{The GRASH\_Rev code}
\label{sect:GRASH_Rev}

\subsection{Gas dynamics}

\begin{figure}[h!]
\begin{center}
\includegraphics[width=0.5\textwidth]{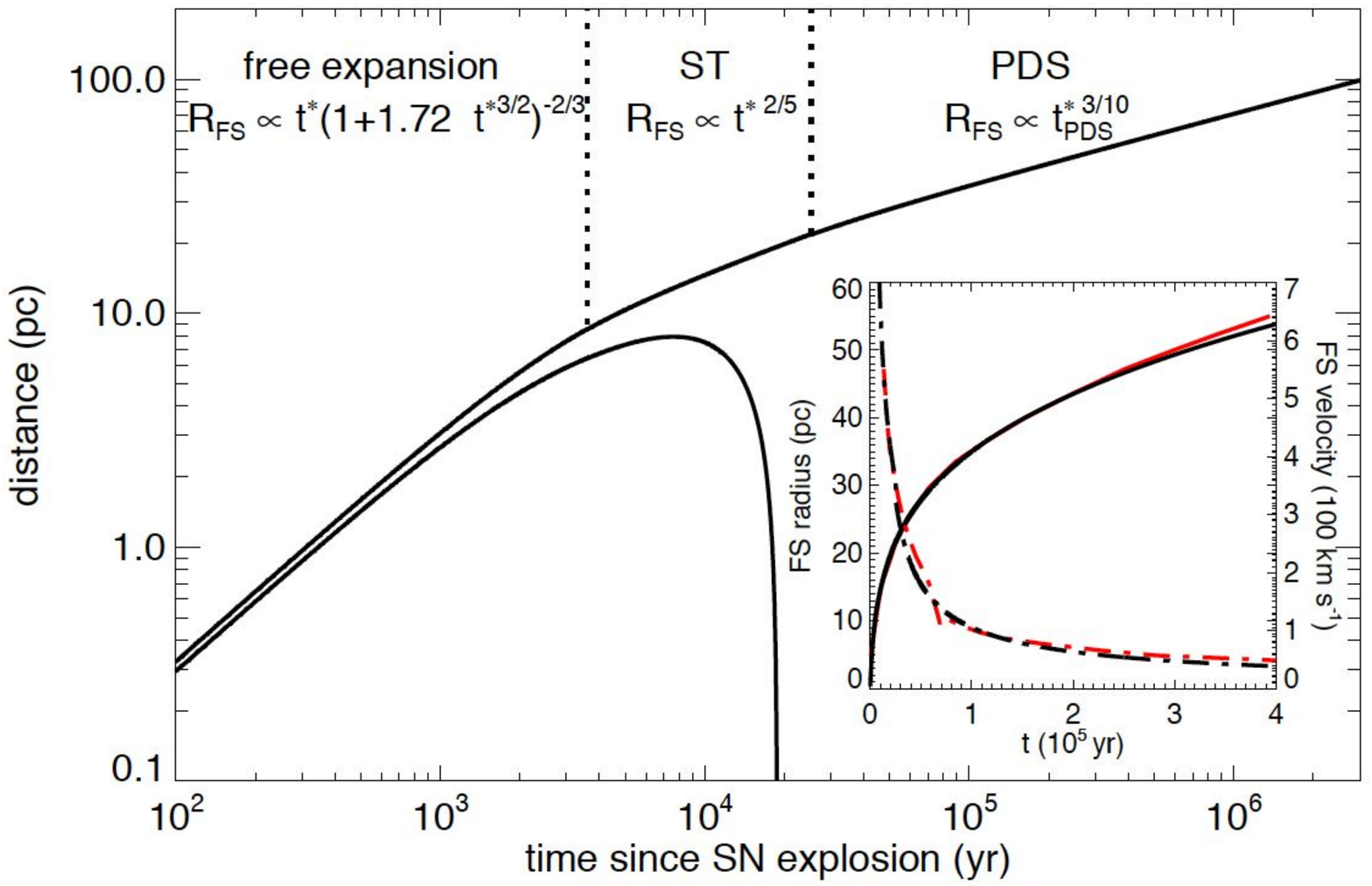}
\caption{Position of the forward and reverse shocks (black lines) as a function of time
from the ED phase through the PDS phase. In the inset we compare the position (solid line) and velocity (dot-dashed lines)
of the forward shock computed using the analytical solution (black lines) with the simulation by \citealt{2015ApJ...803....7S} (red lines).}
\label{fig:SN_evolution}
\end{center}
\end{figure}

The dynamics of the expansion of a supernova is classically divided into three main stages (see Fig.~\ref{fig:SN_evolution}):
\begin{enumerate}
\item{\it Ejecta-dominated (ED) expansion phase:} as the supernova explodes a shock wave is triggered and propagates into the ISM (forward shock, FS).
At the same time a reverse shock (RS) propagates into the ejecta.
The ejecta is in free expansion.
\item {\it Sedov-Taylor (ST) phase:} the mass of the interstellar material swept up by the shock wave becomes comparable 
to the mass of the ejecta. The ejecta is in adiabatic expansion and the radiative cooling is 
negligible (\citealt{1950RSPSA.201..175T,Sedov}).
\item {\it Pressure-driven snowplough (PDS) phase:} the shock wave slows down to velocities $\lesssim 200$\,km/s and radiative cooling 
becomes the main cooling process.
The outermost shell of the ejecta is pushed into the ambient medium due to the
pressure of the hot gas inside it.
\end{enumerate}
Finally, when the velocity of the shock front becomes low enough ($\sim 10$\,km/s), the remnant 
merges with the ISM.

A unified and self-similar solution for the evolution of a SNR from the ED phase
through the ST phase was found by \cite{1999ApJS..120..299T} by comparing
with the results of a hydrodynamical non-radiative simulation.
They considered the following characteristic scales of length and time:
\begin{align}
R_{\rm ch} &\equiv M_{\rm ej}^{1/3}\rho_0^{-1/3}, \nonumber \\
t_{\rm ch}  &\equiv E_{\rm ex}^{-1/2} M_{\rm ej}^{5/6} \rho_{0}^{-1/3},
\end{align}
where $M_{\rm ej}$ is the mass of the ejecta, $\rho_0$ is the ISM density and $E_{\rm ex}$ is the explosion
energy.
The position of the forward ($R_{\rm FS}$) and reverse ($R_{\rm RS}$) shocks with respect to the centre of the 
ejecta for uniform ejecta are well described by:
\begin{align}
R_{\rm FS}(t^{\ast}) &= 2.01R_{\rm ch}t^{\ast}(1+1.72t^{\ast 3/2})^{-2/3}, \nonumber \\
R_{\rm RS}(t^{\ast}) &= 1.83R_{\rm ch}t^{\ast}(1+3.26t^{\ast 3/2})^{-2/3},
\end{align}
during the ED phase, 
while during the ST phase these two quantities can be approximated by:
\begin{align}
R_{\rm FS}(t^{\ast}) &= R_{\rm ch}(1.42t^{\ast} -0.254)^{2/5}, \nonumber\\
R_{\rm RS}(t^{\ast}) &= R_{\rm ch}t^{\ast}(0.779 - 0.106t^{\ast}-0.533 \ln t^{\ast}),
\end{align}
where $t^{\ast} = t / t_{\rm ch}$.

When the forward shock reaches velocities $\lesssim 200$\,km/s these equations 
do not represent a good approximation anymore and radiative cooling must be
taken into account.
\cite{1988ApJ...334..252C} found the fiducial time for the onset 
of the PDS phase to be,
\begin{equation}
t_{\rm PDS} = 1.33 \times 10^4  \frac{E_{51}^{3/14}}{\zeta_{\rm m}^{5/14}n_0^{4/7}}\,{\rm yr},
\end{equation}
\noindent
where $E_{51} = E_{\rm ex} / 10^{51}\,{\rm erg}$, $\zeta_{\rm m}=1$ for solar abundances and 
$n_0$ is the ISM number density.
They found also that the forward shock can be approximated with the following analytical expression:
\begin{equation}
R_{\rm FS} = R_{\rm PDS}\left(\frac{4}{3}t_{\rm PDS}^{\ast}-\frac{1}{3}\right)^{3/10},
\end{equation}
\noindent
where $t_{\rm PDS}^{\ast} = t / t_{\rm PDS}$ and,
\begin{equation}
R_{\rm PDS} = 14 \frac{E_{51}^{2/7}}{\zeta_{\rm m}^{1/7}n_0^{3/7}}\,{\rm pc}
\end{equation}
\noindent
is the position of the forward shock at the time $t_{\rm PDS}$.


The validity of these analytical solutions is guaranteed by the comparison with 
hydrodynamical simulations by \cite{1988ApJ...334..252C} and \cite{1999ApJS..120..299T}.
As a further check, we compare the
analytical approximations with recent hydrodynamical simulations \citep{2015ApJ...803....7S}.
In the inset in Fig.~\ref{fig:SN_evolution} we show the analytical forward shock position and velocity 
(black lines) and those retrieved from Fig. 2 in \citealt{2015ApJ...803....7S} (red lines).
The analytical solution is a good approximation throughout the considered time interval and
the discrepancy in the position of the forward shock is never larger than $\sim 2$\%.

\subsection{Simulation grid dynamics}


We divide the ejecta  into $N_{\rm s}$ spherical shells 
and we assume that they all have the same width $\Delta R = R_{\rm ej} / N_{\rm s}$, where 
$R_{\rm ej}$ is the ejecta radius.
The velocity of each shell is initially determined by the homologous expansion
of the ejecta:
\begin{equation}
v_{j} = v_{\rm ej} \frac{R_{j}}{R_{\rm ej}}, \hspace{1cm} v_{\rm ej} = \sqrt{\frac{10 E_{\rm ex}}{3 M_{\rm ej}}}.
\end{equation}
\noindent
where $v_{j}$ is the velocity of the $j$-th shell and $v_{\rm ej}$ is the velocity of the ejecta radius.
Additional $N_{\rm s}$ spherical shells are defined to model the ISM around the expanding ejecta.
We consider a maximum distance of $R_{\rm max} \sim 200$\,pc from the centre of the ejecta and 
divide the space between $R_{\rm ej}$ and $R_{\rm max}$ in $N_{\rm s}$ shells of equal width 
$\Delta R_{\rm ISM} = (R_{\rm max} - R_{\rm ej}) / N_{\rm s}$.
At the beginning of the simulation, these ISM shells are at rest. 

We then let the system evolve.
The simulation time step, $\Delta t$, is uniquely determined by the
dynamics of the reverse shock. It is defined as the time interval during which 
the reverse shock crosses one shell inside the ejecta, until it reaches the centre of it.
At each given time, shells that have not yet been crossed by the reverse shock
keep expanding homologously and adiabatically. This implies
that the external and internal boundaries of each
shell expand at their initial velocity, leading to a larger volume and to a lower density.
The gas temperature in the shell changes according to:
\begin{equation}
T'_j = T_j \left(\frac{V_j}{V'_j}\right)^{1/\gamma},
\end{equation} 
\noindent
where $T_j$ and $V_j$ are the temperature and volume of the $j$-th shell at a given time $t$, the primed variables
are the same quantities at time $t + \Delta t$ and $\gamma$ is the adiabatic index (assumed to be $\gamma_{\rm SN} = 1.41$, 
\citealt{2013ApJ...764...21C}, for the
gas in the ejecta and $\gamma_{\rm ISM} = 5/3$ for the ISM).
Similarly, shells in the ISM which have not yet been crossed by the forward shock remain at rest
and their gas properties remain unchanged.

On the contrary, when a  shell is invested by the reverse or forward shock, it experiences strong variations
in density, velocity and temperature that are well described by the standard Rankine-Hugoniot jump conditions:
\begin{align}
\label{eq:jump}
\rho_j' &= \frac{\gamma+1}{\gamma-1}\rho_j,\nonumber\\
v_j' &=v_j - \frac{2}{\gamma+1} v_{\rm sh},\\
T_j' &= 2\frac{\gamma-1}{(\gamma+1)^2}\frac{m}{k_{\rm B}}v_{\rm sh}^2\nonumber
\end{align}
where $\rho_j'$, $v_j'$ and $T_j'$ are the density, velocity and temperature
of the $j$-th shell (being it in the ejecta or in the ISM) after the passage of the shock,
$m$ is the mean particle mass, $k_{\rm B}$ the Boltzmann's constant and
$v_{\rm sh}$ is the velocity of the shock in the reference frame of the shell before being
invested by the shock.
This velocity corresponds to $\tilde{v}_{\rm RS}$ and $v_{\rm FS}$ for the reverse and forward shocks,
whose analytical expressions can be found in Truelove \& McKee (1999).

The velocity of shells between the reverse and forward shocks (hit by a shock at earlier times)
is obtained interpolating between the velocity of the reverse and forward shock.
As in \cite{2007MNRAS.378..973B}, we have compared the dynamical evolution predicted by the model with the output of
numerical simulations by \cite{2001A&A...380..309V} and \cite{2003A&A...400..397D},
finding good agreement until $t\sim 10^4$\,yr.

The reverse shock velocity in the reference frame of the unshocked ISM 
reaches an inversion point at $t = t_{\rm rev}$ and then it achieves negative 
velocities, causing the implosion of part of the ejecta.
This process has been investigated in detail through numerical simulations by \cite{1988ApJ...334..252C}.
Before reaching the centre of the ejecta, the temperature and pressure in the core of
the ejecta increase, effectively reflecting the reverse shock, and leading to the 
so-called ``echoes of thunder'' described in their work.
In order to qualitatively reproduce this mechanism we make the following approximation:
for all the shells of the ejecta that are invested by the reverse shock after $t_{\rm rev}$ ($j < j^{\ast}$ with
$j^{\ast}$ the shell invested by the reverse shock at $t_{\rm rev}$),
we set the shell velocity to zero, therefore assuming that the bouncing of the reverse shock 
would effectively cancel out the inward velocity.
The velocity of these shells will remain zero until the end of the simulation and 
shells in between the $j^{\ast}$-th shell and the forward shock have velocities obtained from 
interpolation.
During this phase, radiative cooling of the gas due to atomic processes starts to be important 
and it is taken into account using the cooling function by \cite{1976ApJ...204..290R}.
The error made on the position of the outermost shell of the ejecta computed by this model
is always smaller than a factor of $\sim 2$ when compared with the results of numerical
simulations \citep{1988ApJ...334..252C}. 
Moreover,  since only a quarter of the ejecta is invested by the
reverse shock after $t_{\rm rev}$, all these approximations affect only $\lesssim 2$\%
of the dust mass.

\subsection{Dust coupling}

The dynamics of dust is governed by its coupling to the gas.
Initially, dust is coupled to the unshocked ejecta and uniformly distributed.
For a given shell, $j$, both gas and dust have the same velocity, $v_j$. 
The passage of the reverse shock in the $j$-th shell of the ejecta will affect  
the dynamics of both gas and dust in that shell.
The gas is instantaneously modified according to Eq.~\ref{eq:jump}  and takes 
a velocity $v_j'$,
while dust grains, having more inertia, decouple from the gas and move
with an initial velocity (see Eq.~\ref{eq:jump}):
\begin{align}
\label{eq:inertia}
|v_{\rm rel}| & =  |v_j' - v_j| \nonumber\\
		  & = \frac{2}{\gamma+1}v_{\rm sh}
\end{align}
\noindent
with respect to the shocked gas.

In the model by BS07 dust grains were kept in the same shell for the whole simulation
without considering their full dynamics. Also, once the reverse shock invested a given shell, the grains in that shell were
assumed to be at constant velocity (given by Eq.~\ref{eq:inertia}) in order to take the grain erosion upper limit, 
without considering the action of drag forces.
However, depending on the gas and dust properties, dust grains
modify their velocity according to the following equation (\citealt{1987ApJ...318..674M}):
\begin{equation}
\frac{dy_{\rm rel}}{dt} = \frac{d\chi}{dt} \frac{y_{\rm rel}}{2\chi} - \frac{F_{\rm D}}{m_{\rm gr}},
\end{equation}
\noindent
where $\chi$ is the shock compression, $F_{\rm D}$ the coupling force and $m_{\rm gr}$ the grain mass.
The first term on the right-hand side is known as {\it betatron acceleration} (\citealt{1978ApJ...225..887C,1978ApJ...226..858S}).
In the presence of a magnetic field, $B$, charged (with charge $|q|$) dust grains
gyrate around magnetic field lines with a trajectory determined by the Larmor radius:
\begin{equation}
r_{\rm L} = \frac{m_{\rm gr} v_{\rm rel, \perp}}{|q| B},
\end{equation}
where $v_{\rm rel, \perp}$ is the component of the relative velocity between gas and dust which is perpendicular
to magnetic field lines.
A rapid increase in the gas density due to shock compression will then lead to an enhancement of 
the magnetic field strength and a consequent (betatron) acceleration and reduction of the Larmor radius.
On the contrary, the coupling term reduces the grain velocity as a result of the
 combined effect of collisional (direct) coupling and of plasma drag 
(\citealt{1979ApJ...231...77D}):
\begin{equation}
F_{\rm D}  = F_{\rm D} ({\rm collisional}) + F_{\rm D} ({\rm plasma}).
\end{equation}
The relevant relations that define plasma drag are reported in Eqs. 3 - 6 of \cite{2014A&A...570A..32B}.

\begin{figure}[h!]
\begin{center}
\includegraphics[width=0.5\textwidth]{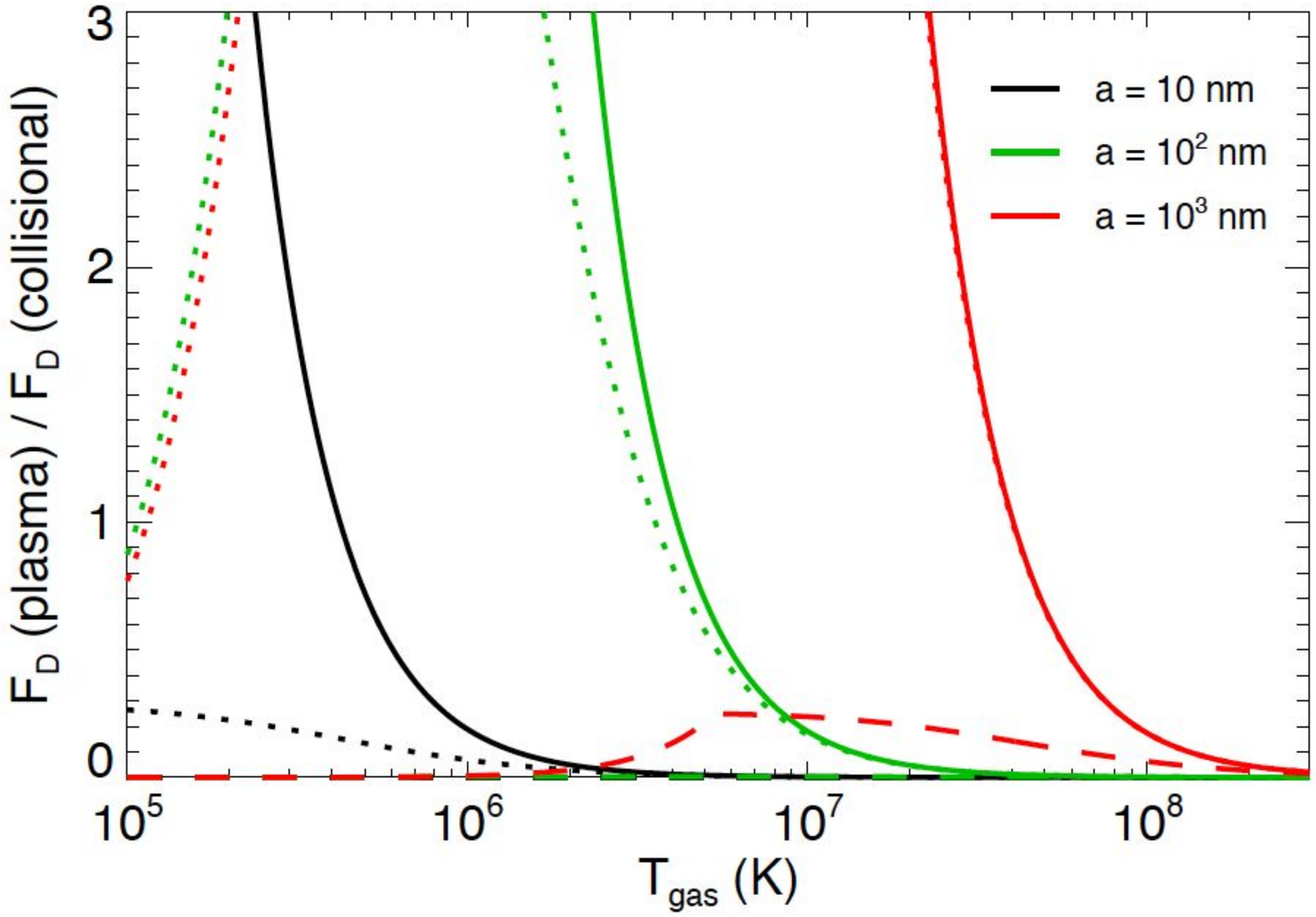}
\caption{Ratio between plasma and collisional drag as a function of the gas temperature for
different grain radii. Drift velocities are indicated by the line style 
($V_{\rm rel} = 10, 100, 1000$\,km/s for solid, dotted and dashed lines, respectively).}
\label{fig:check_plasma}
\end{center}
\end{figure}

In this model we do not describe the ionisation state of the gas.
However, for a sufficiently high gas temperature we can consider the gas  
as fully ionised. Therefore the shocked gas is assumed to be composed by
 H$^+$, He$^{++}$ and O$^+$, which represent the most abundant species.
In Fig.~\ref{fig:check_plasma} we show the ratio between plasma and collisional drag as a function
of the gas temperature and for different grain sizes and drift velocities.
Plasma drag has a negligible effect on the dynamics of small grains, while it exceeds collisional drag
for $\sim 100$~nm grains at gas temperatures $\lesssim 5\times 10^6$\,K.
However, if the velocity is sufficiently large ($v_{\rm rel} \sim 1000~\rm{km/s}$) plasma
drag is negligible at all gas temperatures and for all grain sizes.

It has been shown that betatron acceleration
efficiently affect the dynamics of interstellar dust grains 
invested by supernova-triggered shock waves (\citealt{1994ApJ...433..797J,1996ApJ...469..740J,2014A&A...570A..32B}).
However, this process is probably not relevant for dust produced in SNe.
There is evidence for magnetic field lines distributed radially 
inside the ejecta both from polarimetry observations (e.g. \citealt{2009MNRAS.394.1307D})
and from numerical simulations (e.g. \citealt{2013ApJ...772L..20I,2014MNRAS.437.2802S}).
In the case of symmetric ejecta, the velocity of grains is mostly oriented radially and 
betatron acceleration is therefore negligible inside the ejecta.
In the ISM, shock compression is important during the PDS phase,
when the shock front is at low velocities ($v_{\rm sh} \lesssim 200$\,km/s) and
the gas cools due to the emission of radiation.
However, dust grains ejected from the interior of the SN are almost at rest 
with the post-shock gas during this phase (see Section~\ref{sect:results}) 
and they are therefore not affected by the betatron acceleration.
Hence we do not include betatron acceleration in this version of the code.

\subsection{Dust grain number matrix}

In order to keep track of all the grains in the simulation we define
a matrix, $\mathfrak{M}$, where -  for each of the $2N_{\rm s}$ shells and for each relative velocity 
between grains and gas in the shell - we store the 
number of grains, their size and composition. 
Initially, the grains occupy the matrix elements corresponding to the shells in the ejecta and 
zero relative velocity.
At each time step, grains are displaced according to the laws described above 
and the matrix elements are updated.
A particular care is required when drag forces are too strong and grain velocities
change too quickly during one simulation time step.
In that case the time step is divided into $n_i = 50$ time intervals (each of duration $\delta t$) and 
the grain velocity, $v_{\rm t+\Delta t}$, and position, $d_{\rm t+\Delta t}$, 
are determined by a sub-grid model following:
\begin{align}
v_{\rm t+\Delta t}  &= v_{\rm t} + \sum_{i=1}^{n_i} \left(\frac{dv}{dt}\right)_i \delta t,\nonumber\\
d_{\rm t+\Delta t} &= d_{\rm t} + \sum_{i=1}^{n_i} v_i  \delta t,
\end{align}
where $v_i$ is the intermediate velocity at each time $t + i\delta t$.
Computing the change in grain radius, $a$, at each of the $n_i$ time intervals would have been 
too time-consuming and we therefore account for the grain erosion in this way:
\begin{equation}
a_{\rm t+\Delta t} = a_{\rm t} + \frac{1}{2}\left(\frac{da}{dt}\biggr\rvert_{v = v_{\rm t}} + \frac{da}{dt}\biggr\rvert_{v = v_{\rm t+\Delta t}}\right), 
\end{equation}
where $\frac{da}{dt}$ is the grain radius derivative calculated as described in Section~\ref{sect:proc} and
the subscript indicates the velocity at which the processing is calculated.
The grain radius generally does not change very rapidly and the division in sub-grid time intervals 
is not necessary.

\subsection{Dust processing}
\label{sect:proc}

We consider four main physical processes:
\begin{enumerate}
\item Sputtering due to the interaction of dust grains with particles in the gas;
\item Sublimation due to collisional heating to high temperatures; 
\item Shattering into smaller grains due to grain-grain collisions;
\item Vapourisation of part of the colliding grains during grain-grain collisions.
\end{enumerate}

The sputtering process can be estimated following the formalism by \cite{1994ApJ...431..321T}  as revisited by \cite{2014A&A...570A..32B}.
The sputtering rate is expressed as:
\begin{equation}
\frac{dN_{\rm sp}}{dt} = 2\pi a^2 \sum_i n_i <Y_i v>,
\end{equation}
where $\frac{dN_{\rm sp}}{dt}$ is the number of sputtered atoms per unit time, $a$ is the grain radius, $n_i$
the number density of the $i$-th ion and
\begin{equation}
<Y_i v> = \int Y_i v f_{\rm skM} (v) dv,
\end{equation}
where $Y_i$ is the sputtering yield (assuming a semi-infinite target) of the $i$-th ion, $v$ is the relative velocity between
grain and gas particle (a combination of the thermal and drift velocities) and
$f_{\rm skM} (v)$ is the skewed Maxwellian distribution (\citealt{1978ApJ...226..858S,vinc_thesis,2014A&A...570A..32B}).
We use the same sputtering yields derived by \cite{1994ApJ...431..321T} and used by BS07
and \cite{2006ApJ...648..435N,2007ApJ...666..955N}.

Energetic ions can have a penetration depth
which is comparable or even larger than the smallest grains considered in this work.
For example a $\sim 1$ keV He$^+$ ion (corresponding to the energy of an ion in a hot gas with $T_{\rm g} \sim 8\times 10^6$\,K) 
has a penetration depth $R_{\rm p} \sim 10$\,nm in graphite and therefore it is likely to cross grains of radius smaller than
$R_{\rm p}$ without leading to any sputtering of the target.
This effect was taken into account in models by \cite{1998ApJ...503..247J} and \cite{2008A&A...492..127S}
and then adopted by \cite{2014A&A...570A..32B} for sputtering of interstellar carbonaceous grains 
in supernova-triggered shocks.
Using a similar approach, here we extend the latter study to simulate this effect for all the target 
materials considered in GRASH\_Rev.
The sputtering yield, $Y_{\rm a}$, of a grain of radius $a$ is then expressed as:
\begin{equation}
Y_{\rm a} = Y_{\infty} f(x),
\end{equation}
where $Y_{\infty}$ is the sputtering yield for a semi-infinite target, $x = a / (0.7 R_p)$ and $f(x)$ is a function
that can be parametrised as:
\begin{equation}
\label{eq:size_eff}
f(x) = 1 + p_1 \exp\left[- \frac{(\ln{x/p_2})^2}{2p_3^2}\right] - p_4 \exp\left[-(p_5 x - p_6)^2\right].
\end{equation}
The function $f(x)$ depends on the target material and the six parameters used are 
reported in Table~\ref{table:size_eff}.
In general this function tends to unity for sufficiently large grains (with respect to $R_{\rm p}$, i.e. $x \to \infty$)
and tends to zero for small grains (i.e. $x \to 0$).
For grain radii comparable to the ion penetration depth, the sputtering yield is enhanced and can
reach values $Y_{\rm a} \sim 4 Y_{\infty}$ for graphitic-type grains.
For a detailed explanation of the method used to retrieve the parameters involved in Eq.~\ref{eq:size_eff},
the reader is referred to Appendix~\ref{app:size_eff}.
Finally, since the mass of (sub-)nanoparticles is negligible for all the SN models considered,
we do not implement any molecule dissociation process (e.g. \citealt{2010A&A...510A..36M,2010A&A...510A..37M}) 
as was done in \cite{2014A&A...570A..32B}.

\begin{table}[H]
\caption{Parameters for the analytical modelling of the size-dependent sputtering.\vspace{.2cm}}
\label{table:size_eff}      
\centering          
\begin{tabular}{l | c c c c c c }
\hline\hline
 & Al$_2$O$_3$ & Fe$_3$O$_4$ & MgSiO$_3$ & Mg$_2$SiO$_4$ & AC & SiO$_2$\\
\hline
p$_1$ &  1.1 &        1.2 &       1.0 & 1.5 & 4.9 & 0.85 \\
p$_2$ & 0.39 & 0.55 & 0.50 & 1.2 & 0.55 & 1.2 \\
p$_3$ & 1.1 & 1.0 & 1.0 & 0.57 & 0.77 & 0.56 \\
p$_4$ & 1.9 & 1.9 & 1.8 & 1.1 & 4.7 & 1.1 \\
p$_5$ & 2.1 & 2.0 & 2.1 & 0.52 & 3.0 & 1.6\\
p$_6$ & 0.78 & 0.80 & 0.76 & 0.37 & 1.2 &0.29\\
\hline
\end{tabular}
\end{table}

Sublimation is implemented with the same method described by BS07,
using the thermal sublimation rates by \cite{1989ApJ...345..230G} and \cite{2002ApJ...569..780D}.
In agreement with BS07, we find this process to provide a negligible contribution to dust destruction for all the SN models.

Grain-grain collisions lead to the formation of craters in the two colliding grains.
The matter ejected from the crater is fragmented or even vapourised
and brought back to the gas phase.
Shattering and vapourisation processes are implemented following equations 
by \cite{1994ApJ...433..797J,1996ApJ...469..740J}.
However, some of the grain materials present in this work were not in the original
papers. We adopt values from the literature for the different grain properties (see Table~\ref{table:shat_vap_par}).
For a detailed description of the computation of the relevant quantities the reader is referred to Appendix~\ref{app:shat_vap}.

Given the low density of dust in the ejecta, grain-grain collisions are rare 
and shattering and vapourisation lead to minor processing with respect to sputtering.
This result can be better understood if we calculate the timescale between grain-grain collisions,  $\Delta t_{\rm coll}$.
Particles with a mean number density $\bar{n}$ and a typical radius $a$, in motion in a 
given volume will have a mean free path: 
\begin{equation}
\lambda = \frac{1}{\pi a^2 \bar{n}},
\end{equation}
and the timescale between two collisions is given by:
\begin{equation}
\Delta t_{\rm coll} = \frac{\lambda}{\bar{v}},
\end{equation}
with $\bar{v}$ the typical relative velocity between particles.
As an example we consider a SN $\sim 10^4$\,yr after the explosion.
The volume occupied by dust is roughly a sphere of radius 10 pc.
Assuming 0.5 M$_\odot$ of silicate grains with an average radius of 50 nm, the mean grain
number density is $\bar{n} \sim 10^{-11}$\,cm$^{-3}$.
Then, considering a mean relative velocity of $\bar{v} \sim 500$\,km/s,
the timescale between grain-grain collisions is $\Delta t_{\rm coll} \sim 8\times 10^5$\,yr,
showing that these events are rare.
Furthermore, with the expansion of the ejecta and the dispersion of grains into the ISM, 
the frequency of collisions is even lower.

At each time step in the simulation the grains in all the cells of the matrix $\mathfrak{M}$
are processed by the four mechanisms described above and the matrix is updated.

\begin{figure}[h!]
\begin{center}
\includegraphics[width=0.5\textwidth]{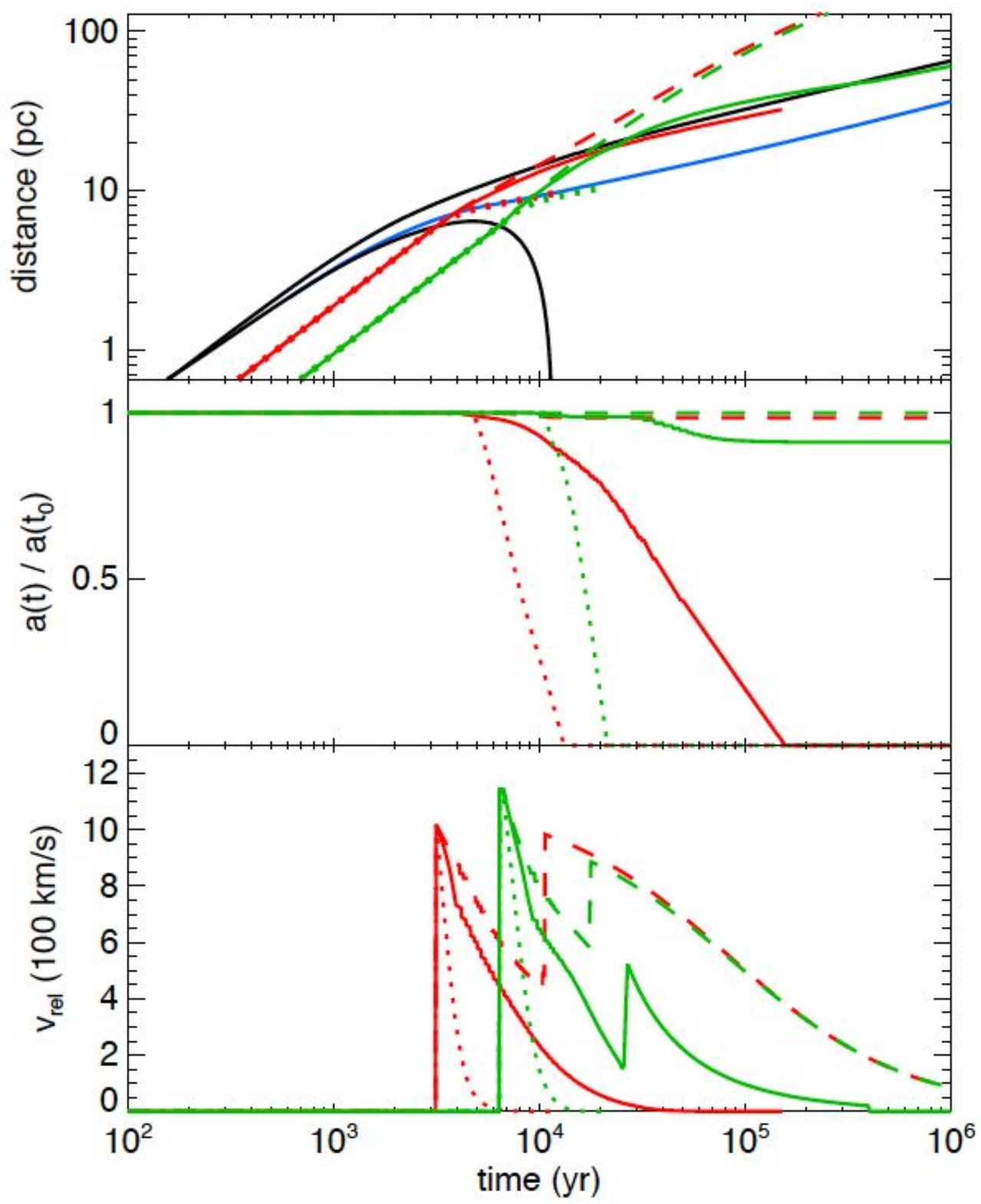}
\caption{Mg$_2$SiO$_4$ test particle trajectory, size and velocity in the ejecta of SNR N49. Green and red
lines indicate particles initially placed at one-forth and half of the radius of the ejecta, respectively. 
The initial size of particles is 10, 100 and 1000 nm for dotted, solid and dashed lines, respectively.
In the top panel, black lines indicate the positions of forward and reverse shock and the blue line is 
the boundary between the ejecta and the ISM.}
\label{fig:testp_plas}
\end{center}
\end{figure}

\section{Results}
\label{sect:results}

We run the GRASH\_Rev code for four SNe with the following input parameters:
the number density of the surrounding ISM, the explosion energy, 
the progenitor mass ($n_{\rm 0}$, $E^{\rm mod}_{\rm exp}$,  $M^{\rm mod}_{\rm prog}$ listed in Table 1)
and the initial dust mass, composition and size distribution ($M_{\rm dust}$, $a_0$, $\sigma$ for each grain species listed in Table 2).
The number of shells used for the simulation grid is fixed and equal to $2N_{\rm s} = 800$.
However, we note that this represents a robust parameter; halving or doubling the number of shells the simulation
output remains unchanged.

In the following sections we present the results of the simulations.

\subsection{Test particle dynamics and processing}

To better understand the behaviour of individual dust grains, we first illustrate the dynamics 
of a few test particles, giving as an input to the code their initial position, size and material.

We consider the simulation of SNR N49 and we define six test particles made of Mg$_2$SiO$_4$ with radii of 10, 100 and 1000 nm initially
placed at a distance of one-forth and half of the initial radius of the ejecta.
In Fig.~\ref{fig:testp_plas} we show their trajectories (upper panel), the size evolution (central panel)
and relative velocities with respect to the gas. 
The position of forward and reverse shocks are indicated by black lines while the position
of the boundary between the ejecta and the ISM by the blue line.

Initially all the particles are at rest with respect to the ejecta.
When the reverse shock reaches the shell where the particles reside, the 
gas undergoes a jump in conditions (Eq.~\ref{eq:jump}) 
while dust grains are not instantaneously
coupled and assume a velocity given by Eq.~\ref{eq:inertia} with respect to the gas.
Immediately after the passage of the reverse shock, depending on the gas conditions and
grain size, dust grains are slowed by drag forces and processed due to collisions with gas particles
and other grains.

Small grains ($a = 10$ nm, dotted lines) are quickly stopped and destroyed within the ejecta,
while larger grains ($a = 100, 1000$ nm, solid and dashed lines, respectively)
are eroded to a lower extent. 
Furthermore, 100 nm grains which were initially placed at one-forth of the 
ejecta (green lines) have enough inertia to cross the forward shock while 
100 nm grains initially placed at half of the ejecta (red lines) are stopped and destroyed 
in $\sim 10^5$\,yr. 

As soon as a grain crosses the forward shock, its relative velocity with respect
to the unshocked ISM gas, $v'_{\rm ISM}$, will be enhanced to
\begin{equation}
v'_{\rm ISM} = v_{\rm ISM} + v_{\rm sh},
\end{equation}
where $v_{\rm ISM}$ is the relative velocity between the grain and the shocked
ISM and $v_{\rm sh}$ is the shock velocity.
Because of the effect of drag forces, grains will again slow down in the unshocked
ISM. In this medium, if the velocity of a grain is too low (e.g. green solid line in Fig.~\ref{fig:testp_plas}), 
it can be reached by the faster forward
shock and will cross the shock front a second time.
This time, the relative velocity between grain and gas is reduced
and the grain is almost at rest in the shocked gas, therefore avoiding further processing.

Hence, it is clear that the fate of individual grains depends not only on its initial size but
also on its initial position within the ejecta.

\subsection{Dust mass evolution}

\begin{figure}[h!]
\begin{center}
\includegraphics[width=0.5\textwidth]{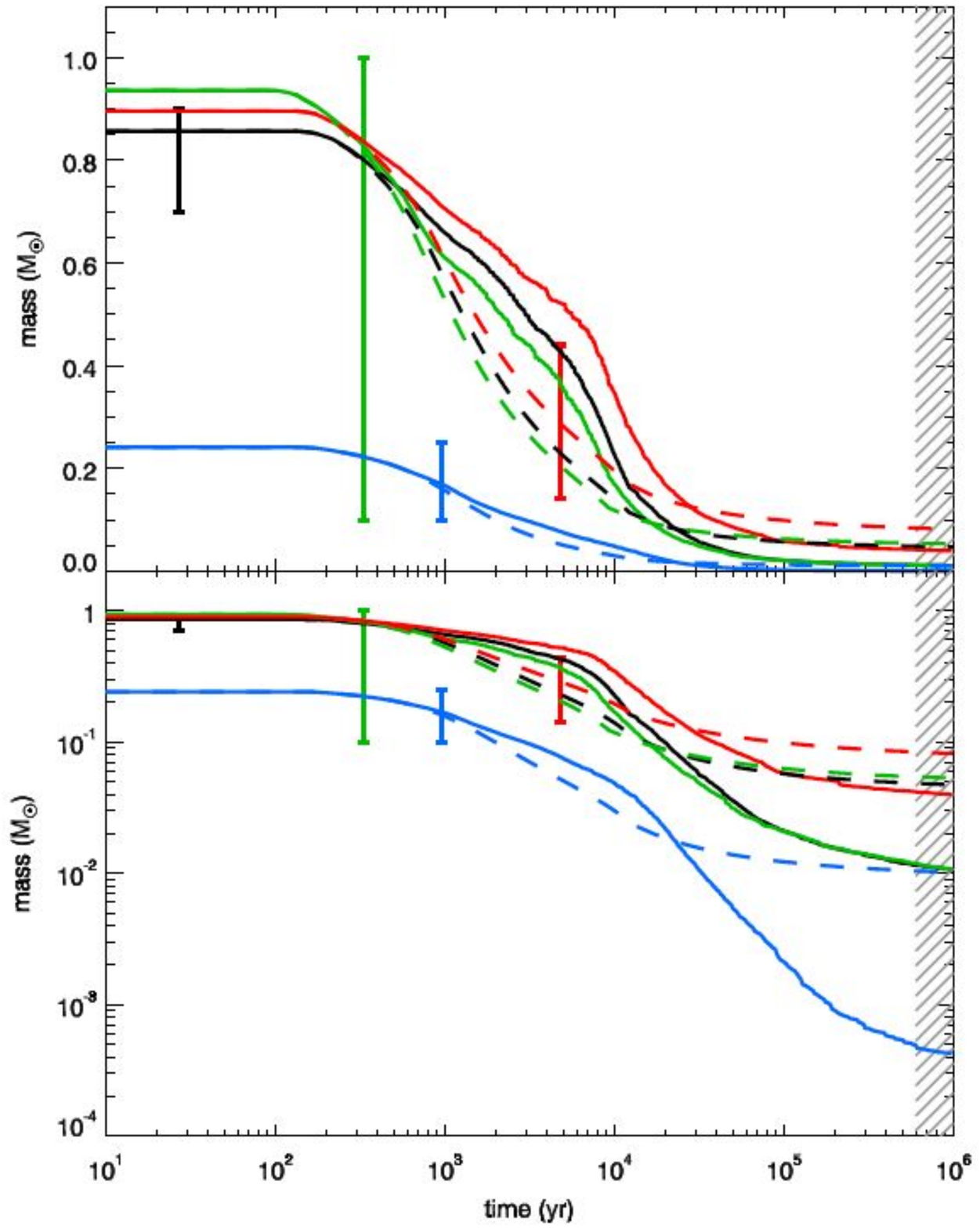}
\caption{Dust mass evolution as a function of time for the four SNe considered in this study (1987A: black, CasA: green,
Crab: blue and N49: red). 
Solid lines represent GRASH\_Rev results and dashed lines the results obtained by BS07 model using the same initial conditions. 
Data points represent the observed dust masses and the shaded region indicate the time interval when dust processing fades out.}
\label{fig:mass_comp}
\end{center}
\end{figure}

In Fig.~\ref{fig:mass_comp} we show the time evolution of the dust mass following
the SN explosion. The evolution starts at the end of the dust nucleation phase.
Each line represents the evolution of a single SN (1987A: black, CasA: green,
Crab: blue and N49: red) and the data points illustrate the
dust mass estimated from the observations (see Section~\ref{sect:sample}).
As a comparison we also show the mass evolution computed using the model by 
BS07, assuming no drag forces (dashed lines).
In the inset we highlight results for the time interval $t  = 10^5 - 10^6$\,yr in a logarithmic scale.

The dust mass estimated using GRASH\_Rev simulations are in good agreement with the
observations, except for SNR N49 for which the dust mass predicted by GRASH\_Rev  is $\sim 15\%$ larger 
than the upper mass limit of the range of values inferred from observations (0.4 $M_{\odot}$, see Table~\ref{table:par_sample} and 
section~\ref{sec:dustmassdistri}).
A more detailed comparison between model predictions and observations is given in Section~\ref{sec:dustmassdistri}.

At times $t \sim 10^3 - 10^4$\,yr the reverse shock destruction predicted by GRASH\_Rev is less important \
than that predicted by BS07. In fact, during this time interval, because of the passage of the
reverse shock dust is decoupled from the gas and travels in between the reverse and forward shocks 
where the gas conditions are milder.  
For $t \gtrsim 10^4$\,yr, most of the grains reach regions behind the forward shock
where gas conditions are harsh and dust erosion is efficient.
For all the four SNe, we predict final dust masses which are
$\sim 5 - 50 \%$ of those estimated using the model by BS07, where the displacement of grains and the effect
of the forward shock were neglected.

\subsection{Dust mass distribution}
\label{sec:dustmassdistri}

\begin{table*}[t!]
\caption{Dust masses (expressed in $M_{\odot}$) at the observation time, $t_{\rm obs}$, and at the end of the simulation, $t_{\rm end}$.\vspace{.2cm}}
\label{table:dust_mass}      
\centering          
\begin{tabular}{@{\extracolsep{1mm}} l c c c c  c c c c }
\hline\hline
  &  \multicolumn{2}{c}{1987A}   & \multicolumn{2}{c}{CasA}  & \multicolumn{2}{c}{Crab}& \multicolumn{2}{c}{N49}\\
\cline{2-3}
\cline{4-5}
\cline{6-7}
\cline{8-9}
  & $t_{\rm obs}$ & $t_{\rm end}$   & $t_{\rm obs}$ & $t_{\rm end}$  & $t_{\rm obs}$ & $t_{\rm end}$   & $t_{\rm obs}$ & $t_{\rm end}$ \\
\hline
$M_{\rm Al_2O_3}$ &     $2.57\times10^{-2}$& $1.24\times 10^{-6}$ & $2.14\times10^{-2}$ & $3.00\times 10^{-6}$  & $3.66\times10^{-3}$ & $1.51\times10^{-6}$ & $6.19\times10^{-3}$    & $3.93\times10^{-6}$ \\
$M_{\rm Fe_3O_4}$ &     0.11                        & $1.46\times 10^{-5}$ &  0.13                         & $6.78\times 10^{-6}$ & $1.58\times 10^{-2}$ & $2.00\times10^{-6}$ & $3.02\times 10^{-2}$  & $2.27\times10^{-5}$\\
$M_{\rm MgSiO_3}$ &     $6.58\times10^{-6}$& $5.91\times 10^{-8}$ & $2.4 \times10^{-5}$ & $1.65\times 10^{-9}$ & $1.80\times 10^{-3}$ &  $1.87\times 10^{-6}$  & $1.84\times10^{-5}$ & 
$1.70\times 10^{-7}$\\
$M_{\rm Mg_2SiO_4}$ & 0.44                         & $2.10\times 10^{-3}$ & 0.40                           & $1.34\times 10^{-5}$ & $6.19\times 10^{-2}$ & $1.38\times10^{-4}$ & 0.28                          & $1.48\times10^{-3}$\\
$M_{\rm AC}$  &         $7.15\times10^{-2}$     & $8.50\times 10^{-3}$ & 0.12                         & $1.07\times 10^{-2}$ & $7.85\times 10^{-2}$ & $2.57\times10^{-4}$ & 0.11                          & $3.86\times10^{-2}$\\
$M_{\rm SiO_2}$ &          0.19                         & $1.26\times 10^{-4}$ & 0.16                         & $6.15\times 10^{-6}$ & $6.80\times10^{-3}$  &  $2.08\times10^{-5}$ & $9.65\times10^{-2}$   & $3.51\times10^{-4}$\\
\hline
$M_{\rm TOT}$  &      0.84                               & $1.07\times 10^{-2}$ & 0.83                        & $1.07\times10^{-2}$  & 0.17                           & $4.21\times10^{-4}$  & 0.52                        & 
$4.02\times 10^{-2}$\\
\hline
\end{tabular}
\end{table*}

\begin{figure}[h!]
\begin{center}
\includegraphics[width=0.5\textwidth]{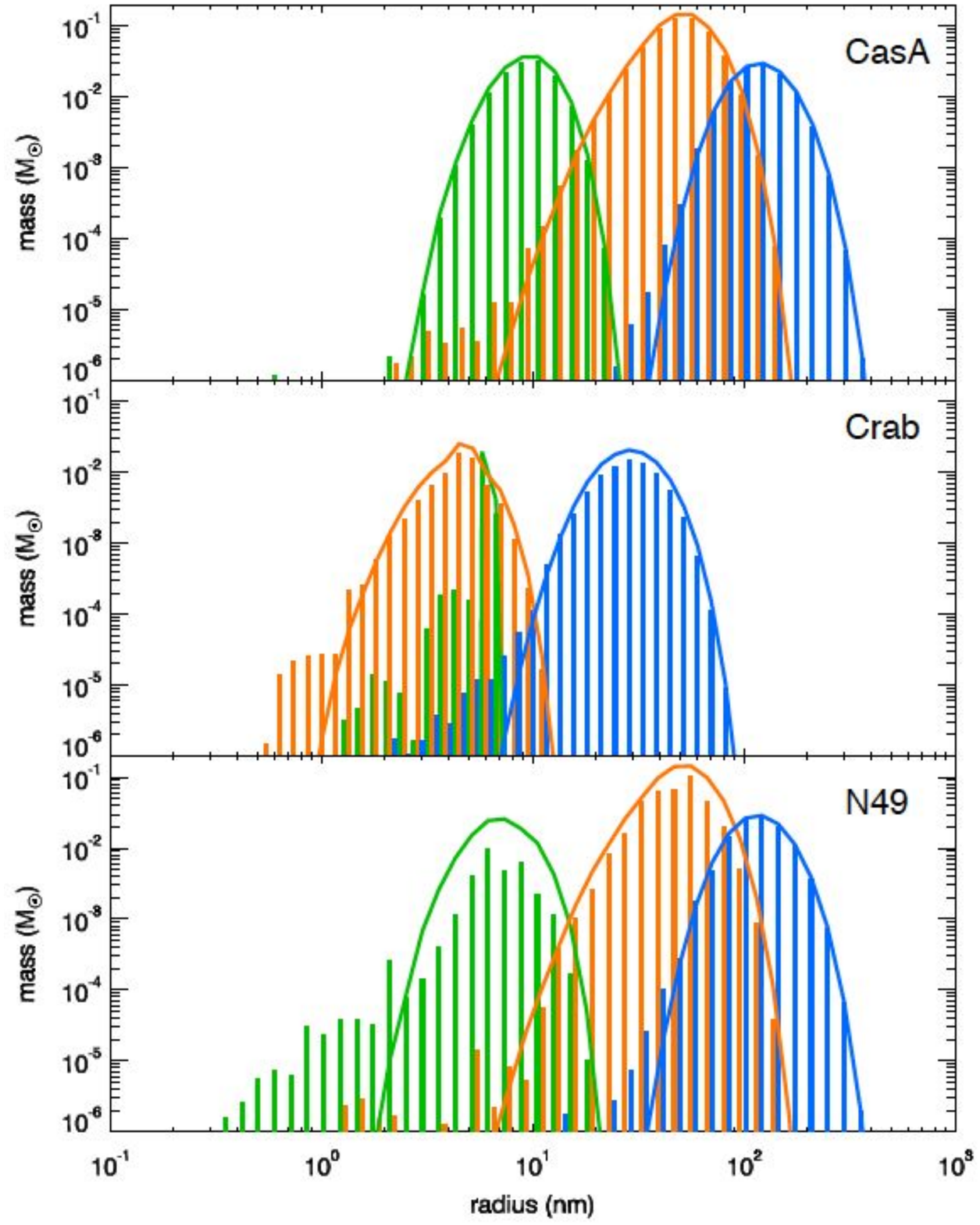}
\caption{Dust mass distributions for CasA, Crab and N49 at formation (solid lines) and 
observation times (histograms).
In each panel we show the three dominant dust species: Fe$_3$O$_4$ (green),
Mg$_2$SiO$_4$ and SiO$_2$ (summed up together in orange) and
AC (blue).}
\label{fig:distfin}
\end{center}
\end{figure}

In Fig.~\ref{fig:distfin} we show the predicted mass in different dust species as a function of the grain size for Cas A, Crab and N49.
The initial mass distributions (obtained as described in Sect.~\ref{sect:structure_ejecta})
are shown with solid lines, while the dust mass distributions at the time
corresponding to the age of these SNe, $t_{\rm obs}$, is represented by the histograms.
Only the most abundant species are displayed here: Fe$_3$O$_4$ (green),
Mg$_2$SiO$_4$ and SiO$_2$ (summed up together in orange) and
AC (blue).We do not show the results for SN 1987A because the reverse shock has not yet affected the mass of dust
formed in this SN (see Fig.~\ref{fig:mass_comp}). 

The dust mass distribution at the observation time is shifted towards 
small grains and reduced in mass as a result of dust processing (compare the entries in Tables 2 and ~\ref{table:dust_mass}).
This effect is more evident for the oldest SN remnant, N49.
Furthermore, small grains are more affected by processing than 
large grains, reflecting the behaviour of test particles seen in Fig.~\ref{fig:testp_plas}.

In what follows, the dust size and composition predicted by GRASH\_Rev for these three SNe
at their respective ages is compared with the dust properties adopted to estimate the dust mass 
from IR and submm observations. 

For CasA our model predicts little destruction and the three dominant species
are characterized by a mass distribution which is similar to the initial one.
About $0.77$ M$_{\odot}$ of dust is still unshocked and therefore only 0.06 M$_{\odot}$ of dust is
likely to be warm and collisionally heated by the hot post-shock gas.
These values are in good agreement with the mass of cold and warm dust obtained from fitting observations  ($0.1 - 1 M_\odot$ from
{\it Herschel} and polarimetry observations and $0.02 - 0.054$ M$_{\odot}$ from {\it Spitzer} data, \citealt{2008ApJ...673..271R,2014ApJ...786...55A}).

The Crab Nebula has been observed with {\it Herschel} at wavelenghts between 51 and 670 $\mu$m by \cite{2012ApJ...760...96G}.
They fit the observed emission with two-temperature blackbodies, inferring a dust mass of 0.11 (0.24) $M_{\odot}$
when carbonaceous (silicate) grain optical properties are used to fit the data.
At the time corresponding to the age of this SN, we predict a total dust mass of
0.17 $M_{\odot}$, in good agreement with the observations, 
with an almost equal contribution from silicates ($\sim 0.09 \, M_\odot$) and carbon grains ($\sim 0.08 \, M_\odot$).
However, the predicted  grain composition is more complex than assumed to fit the data.
%
%
The simulation predicts that  $\sim 0.023 \, M_\odot$ of dust lies between the forward and reverse shock,
probably emitting at high temperatures, while the warm dust mass inferred from the data
is estimated to be $6- 8 \times 10^{-3}\,M_{\odot}$. However, in order to fit the data, all grains are assumed 
to be either at high temperature ($T_{\rm w} \sim$ 55 - 63 K) 
or at low temperature ($T_{\rm c} \sim$ 28 - 33 K). Given the large range of 
predicted grain sizes, this assumption is likely to be a very poor approximation.
%
%

Observations of warm dust emission in N49 by \cite{2010A&A...518L.139O} indicate a dust mass of 0.1 (0.4) $M_{\odot}$
if carbonaceous (silicate) grains are assumed as the main constituent of dust.
The cold component is hard to detect observationally as it is strongly contaminated by the emission of dust
in the parent cloud at a similar temperature. Only a single dust temperature has been used to fit the data
 \citep{2010A&A...518L.139O} and this could lead to a large uncertainty in the estimated dust mass \citep{2015MNRAS.449.4079M}.
GRASH\_Rev simulation predicts that in SNR N49 the reverse shock has already processed much of the ejecta and only 0.04 $M_{\odot}$
of dust is still in the unshocked region, emitting as cold dust.  Hence, we predict a warm dust mass of $\sim 0.48 \, M_\odot$,
dominated by silicate grains, which is only $\sim 15\%$ larger than the observed warm dust mass when silicate optical properties
are assumed.

Given the uncertainties in inferring the dust mass from the data, we can conclude that the model predictions 
provide a reasonable estimate of the mass of dust that has survived the passage of the reverse shock. Based on our simulation, we can
conclude that a dust mass of $[0.7 - 0.9]~M_\odot$, such as the one observed in SN 1987A, can be safely considered to be representative of the efficiency of
dust production in massive stars, as the ejecta have not yet been invested by the reverse shock. This conclusion does not apply to the other three SNe
that we have considered, where - at the estimated time from the explosions - between 10 and 40\% of the freshly formed dust has already been destroyed.
However, for none of the SNe considered, the reverse shock travelled to the center of the ejecta. This implies that current observations can not be used
to estimate the mass of dust that will be injected in the interstellar medium, the so called {\it effective dust yield}.

\subsection{The effective SN dust yields}

The final time of the simulation is assumed to be $t_{\rm end} = 10^6$\,yr where the velocity of the forward shock is 
$v_{\rm sh} \lesssim 10$\,km/s and dust grains are either destroyed or embedded in the ISM gas.
Table~\ref{table:dust_mass} illustrates the predicted dust mass at the end of the simulation. 
We find that the effective dust masses injected in the ISM from these SNe will be significantly 
smaller than inferred from present observations:  $\sim 1 \%$ of the currently detected mass for 
SN1987A and CasA, $\sim 8 \%$ for N49 and less than $1\%$ for the Crab nebula. The resulting dust yields 
depend on the progenitor mass, explosion energy and the age of the SN, and lie in 
the range $[4.21 \times  10^{-4} - 4.02 \times 10^{-2}]\, M_\odot$.
Grains in the Crab nebula suffer a largest destruction. 
In fact, the Crab nebula has a low explosion energy and progenitor mass compared to the other SNe.
This leads to the formation of smaller dust grains (e.g. AC grains have an average 
size of $a_0 = 26$\,nm while in the other three SNe $a_0 \sim 90-100$\,nm) and 
to a lower dust mass. 
For the same reason, silicate grains are destroyed more efficiently. As a result,
more than 60\% of the dust mass that will be injected in the ISM is in the form of carbonaceous grains. 
These grains, being larger than grains of other species, have more inertia and, depending on their initial
position, can cross both the reverse and forward shock, reaching the unshocked ISM and surviving the explosion.

\section{Discussion and astrophysical implications}
\label{sect:discussion}

The dust mass obtained from fitting observations is determined 
under several assumptions, which make the measurement rather uncertain.
\begin{enumerate}
\item Often, a single dust material and size is used to reproduce the dust emission.
However, dust in SNRs is rich in composition and each material has a wide size distribution.
This is reflected in the dust temperature distribution: for example small grains are heated to 
higher temperatures compared to larger grains and their emission per unit mass will be higher.
\cite{2015MNRAS.449.4079M} noted that depending on the grain size/temperature distribution 
the measured dust mass can vary up to six times the original value but that most of the 
complication derives from the assumed grain material.
\item In many cases and especially in young SNe, most of the dust mass resides in the
unshocked ejecta and is therefore likely to be cold with a similar temperature to that of dust in
the parent cloud (radiative transfer calculations in SN1987A predict temperatures $T_{\rm dust}=15-20$\,K; \citealt{2015MNRAS.446.2089W}).
The emission from these two sources comes from the same spatial region and are therefore
entangled and further multi-wavelength observations are therefore needed for a correct
dust mass evaluation.
\item For CasA and the Crab Nebula, recent observations suggest that 
the unshocked ejecta have a clumpy density distribution (\citealt{2015arXiv150705399L,2015ApJ...801..141O}).
When this is taken into account, the inferred dust masses are generally larger. 
In dust formation/destruction models,  a clumpy ejecta
would also favour a larger dust mass as ({\it i}) the formation of dust grains would be more efficient, therefore creating larger grains and
({\it ii}) the higher contrast in density between the diffuse and clumpy gas would slow 
the reverse shock, leading to less dust processing.
\end{enumerate}

In addition, the mass of the progenitor stars and the explosion energy for some SNe are also difficult
to reconstruct from the available data. These do not impact the estimated mass of freshly
formed dust\footnote{As an example, changing the explosion energy
adopted for CasA from $1.5 \times 10^{51}$erg to $3 \times 10^{51}$erg induces only a ~ 5 \% difference
in the dust mass. A similar difference is found if we increase the Ni$^{56}$ mass adopted for the Crab
from $0.014 M_\odot$ to 0.058 $M_\odot$.} but may affect the destruction efficiencies by the 
reverse shock.  
As an example, assuming an increase in the explosion energy
by a factor 2 for Cas A leads to 60\% smaller dust mass at $t_{\rm obs}$ than the value reported in Table 4.
However, as it is clear from Figure 4, this model would still be in agreement with the
observed dust mass, given the large errorbars. 
Hence, the uncertainties which affect the measured dust mass
encompass the range of dust masses predicted by the model for different initial explosion parameters.

Given these uncertainties, the models described in this paper 
appear to well reproduce the dust masses in the four SNe that we have considered.
The average effective dust yield is estimated to be $(1.55 \pm 1.48) \times 10^{-2}\,M_{\odot}$.
This is in good agreement with the dust yield used in chemical evolution models
based on BS07 dust yields with moderate destruction by the reverse shock. For a Salpeter initial mass function (IMF),
 the average dust mass injected by SNe in the ISM is $(2.50 - 2.73)\times 10^{-2} M_{\odot}$
for SN progenitors with metallicities $0.1 \, Z_\odot  \le Z \le 1 \, Z_\odot$ and a circumstellar medium 
density of 1 cm$^{-3}$ (\citealt{2009MNRAS.397.1661V,2011MNRAS.416.1916V,2014MNRAS.444.2442V}; see also 
Fig.~6 in \citealt{2014MNRAS.442.1440S}).

In the Milky Way, the timescale between the explosion of two SNe has been estimated to be
$\sim 40$\,yr (\citealt{2011MNRAS.412.1473L}). Using the mean effective dust yield that we 
have obtained leads to a SN dust production rate of $(3.9 \pm  3.7) \times 10^{-4}\,M_{\odot}$\,yr$^{-1}$.
This value can be compared to the dust production rate by AGB stars. Because of
source confusion in the Galactic plane, surveys have targeted the Magellanic Clouds to make a global
census of dusty AGB stars, inferring dust production rates of $\sim 2 \times 10^{-5} \, M_\odot$/yr for
the Large Magellanic Cloud (LMC, \citealt{2012ApJ...753...71R}) and of $\sim 8 \times 10^{-7} \, M_\odot$/yr for the
Small Magellanic Cloud (SMC, \citealt{2012ApJ...748...40B}), with variations among
the results of different surveys (\citealt{2013MNRAS.429.2527M}), and in good agreement with
theoretical predictions  (\citealt{2012MNRAS.420.1442V,2012MNRAS.424.2345V,2014MNRAS.439..977V}).
Hence, even in the LMC, the current dust production
rate by AGB stars appears to be more than one order of magnitude smaller than that estimated from our sample of four SNe. 
Yet, the relative importance of AGB stars and SNe as dust producers depends on the initial mass function and on 
the star formation history of each galaxy (\citealt{2009MNRAS.397.1661V}). 

Yet, the importance of stellar sources of dust is limited by grain destruction in SN-driven interstellar shocks. 
The dust destruction timescale in the Milky Way has been estimated to be $\tau_{\rm des, MW} \sim 10^8, 10^9$\,yr by \cite{2014A&A...570A..32B}
and \cite{2015ApJ...803....7S}, respectively.
Rescaling these values by the ratio between the ISM mass in the Milky Way and in the LMC ($M_{\rm ISM, MW} / M_{\rm ISM, LMC} \approx 100$),
the dust lifetime in the LMC can be estimated as $\sim 10^6 - 10^7$\,yr, in agreement with the recent observational study by \cite{2015ApJ...799...50L},
which leads to $\tau_{\rm des, LMC} \sim 2\times 10^7$\,yr.
Then, assuming a total interstellar dust mass in the LMC of $M_{\rm dust} = 7.3 \times 10^5\,M_{\odot}$ (\citealt{2014ApJ...797...85G}),
the resulting dust destruction rate is $dM_{\rm des, LMC} / dt \sim 3 \times 10^{-2} - 7 \times 10^{-1}\,M_{\odot}\,{\rm yr}^{-1}$.
These values are larger than the inferred dust production rates for SNe and AGB stars by a factor of $> 100$, therefore requiring 
dust accretion in the gas phase of the LMC.
Even if the dust formation timescale in AGB stars and SNe appears to be longer than the
destruction timescale, stellar dust production is nevertheless an important mechanism 
to provide the seeds for grain growth via gas accretion in denser regions (see however \citealt{2011A&A...530A..44J}).

The result of our study suggests that - among all the different dust species that are formed in the ejecta of
SNe - carbonaceous grains are those which suffer the least destruction by the reverse shock, providing the
dominant contribution to the effective SN dust yields. 
This implies that carbon dust may be present even in the ISM of young galaxies, 
with relatively unevolved stellar populations.
This selection effect is even more pronounced if carbon grains are preferentially formed
with densities lower than graphite ($\rho < 2.2$\,g cm$^{-3}$, see Appendix~\ref{app:carb_dens}).

Finally, it is not straightforward to use the results of this study as a way to estimate the mass of dust produced
by SNe in galaxies at high redshifts. In fact, although very little is known about their ISM conditions,
if these SNe explode in a circumstellar medium of larger density, they experience a harder reverse shock
and presumably a stronger dust destruction. As an example, if exploded in a circumstellar
medium density of $n_0 \sim 15$\,cm$^{-3}$ - a factor 10 larger than assumed here - 
the same simulation used to describe SN 1987A predicts a final dust mass of $3.4\times 10^{-4}$\,M$_{\odot}$,
which is $\sim 30$ smaller than our standard run.

\section{Conclusions}
\label{sect:conclusions}

We have presented GRASH\_Rev, a new code which describes dust processing in a SN explosion,
improving the existing BS07 model by treating the full dynamics of dust grains within the ejecta
and in the surrounding ISM.  We have applied the model to 4 well studied SNe in our Galaxy
and in the LMC: SN 1987A, Cas A, the Crab Nebula and N49. Multi-wavelength observations
with {\it Spitzer} and {\it Herschel} allow to infer the mass of warm and cold dust associated
to the ejecta. 

We find that our models are able to well reproduce the observed dust mass and can therefore be
used to estimate the effective dust yield, hence the mass of dust produced by SNe that is able to
enrich the ISM.

For SN 1987A, the ejecta have not yet been reprocessed by the reverse shock. Hence, the 
currently observed dust mass of $[0.7 - 0.9]\, M_\odot$ is indicative of the efficiency of dust
formation in SN ejecta. Conversely, between $10$ and $40\,\%$ of the newly formed dust has
already been destroyed in the other three SNe. The simulations show that sputtering
due to the interaction of dust grains with particles in the gas represents the dominant 
destruction process because physical conditions in the ejecta cause 
sublimation to be almost negligible and grain-grain collisions to be extremely rare. 

Since the largest dust destruction occurs between $10^3$ and $10^5$~yr after the explosion, 
current observations which sample relatively young SNe (with age $< 4800$\,yr) 
provide only an upper limit on the effective dust yield. Our models indicate that only between 1 and 8\%
of the currently observed dust mass will contribute to the enrichment of the ISM, with an average
SN effective dust yield of $(1.55 \pm 1.48) \times 10^{-2}\,M_{\odot}$, largely dominated by carbonaceous grains. 
This is in good agreement with the IMF-averaged yields adopted in chemical evolution models 
(\citealt{2009MNRAS.397.1661V,2011MNRAS.416.1916V,2014MNRAS.444.2442V,2014MNRAS.445.3039D,2014MNRAS.442.1440S}) 
which account for SN reverse shock destruction. When compared to dust destruction efficiencies in SN-driven
interstellar shocks recently estimated by theoretical models (\citealt{2014A&A...570A..32B}; \citealt{2015ApJ...803....7S}) and observations \citep{2015ApJ...799...50L}, these figures imply that SNe
may be net dust destroyers, pointing to grain growth in the ISM as the dominant dust enrichment 
process both in local galaxies and at high redshifts (\citealt{2014MNRAS.444.2442V,2014MNRAS.445.3039D,2015A&A...577A..80M,2015MNRAS.451L..70M}).

\begin{acknowledgements}
The research leading to these results has received funding from the 
European Research Council under the European Union (FP/2007-2013) / ERC
Grant Agreement n. 306476.
\end{acknowledgements}

\bibliographystyle{template/aa}
\bibliography{bib_GRASHRev}{}

\appendix

\section{Carbonaceous grain density}
\label{app:carb_dens}

Carbonaceous grains formed in SNe exhibit a large range of density (1.6 to 2.2 g cm$^{-3}$) as it is revealed by 
various measurements from presolar grains (e.g. \citealt{2005LPI....36.1867A,2014AIPC.1594..307A}).
In our work we assumed graphitic-type grains with a density of 2.2 g cm$^{-3}$.
In this appendix we consider low-density ($\rho = 1.6$\,g cm$^{-3}$) amorphous carbon grains 
and we show the impact of this change on the amount of carbonaceous dust surviving the explosion of a SN.

\begin{figure}[h!]
\begin{center}
\includegraphics[width=0.5\textwidth]{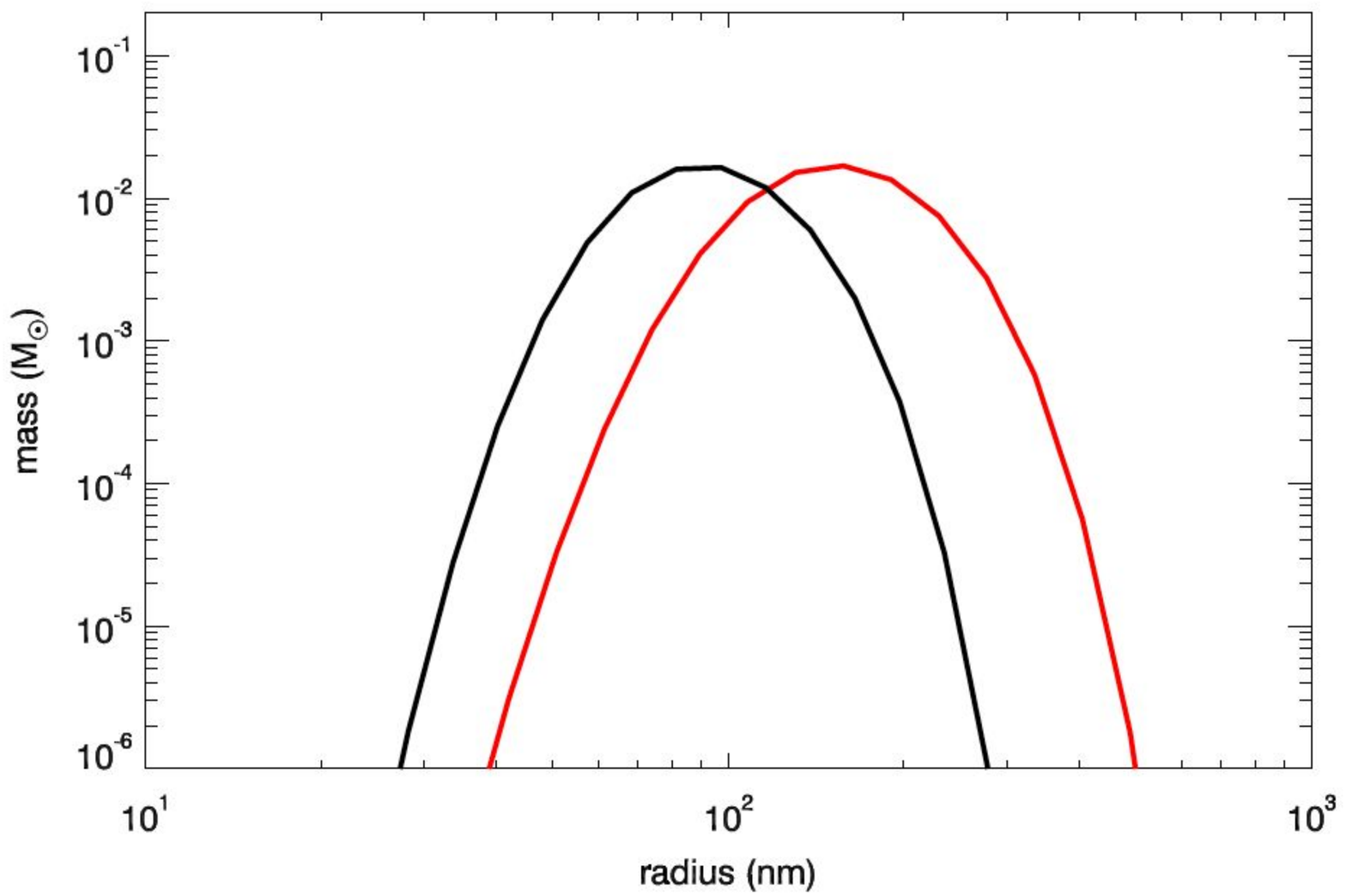}
\caption{Dust mass distribution assuming graphitic-type ($\rho = 2.2$\,g cm$^{-3}$, black line) or amorphous carbon ($\rho = 1.6$\,g cm$^{-3}$, red line) grains.}
\label{fig:dist_graph_AC}
\end{center}
\end{figure}

We consider as an example the case of SN 1987A and rerun the dust formation model lowering the density of carbonaceous 
grains.
In Fig.~\ref{fig:dist_graph_AC} we show the initial dust mass distribution for graphitic-type ($\rho = 2.2$\,g cm$^{-3}$, black line)
and for low-density amorphous carbon grains ($\rho = 1.6$\,g cm$^{-3}$, red line).
While the total carbonaceous dust mass formed in the two cases is very similar ($M_{\rm TOT} = 7.15 \times 10^{-2} M_\odot$ and $7.02 \times 10^{-2} M_\odot$
for high- and low-density carbon grains, respectively), the mass distribution in the case of a lower density
is shifted towards larger grains ($a_0$ is shifted from $\sim 90$\,nm to $\sim 160$\,nm in the case of low-density grains).

We then updated the GRASH\_Rev code taking into account the carbonaceous grain processing described by \cite{2014A&A...570A..32B}. 
In particular, the lower density of the material shows its biggest impact on the sputtering yield, which is enhanced by up to an order
of magnitude over a large range of energies and grain sizes (see Figs.~3 and 4 in \citealt{2014A&A...570A..32B}).
However, the mass distribution of low-density carbon grains is shifted towards larger grains which are therefore
less affected by sputtering than smaller, higher-density grains.
This cancels out the enhancement of the sputtering yield.
The resulting mass of carbon dust surviving the explosion of SN 1987A is then enhanced by a factor $\sim 3$ from
$M_{\rm AC} = 8.5 \times 10^{-3}\,M_\odot$ to $2.5\times 10^{-2}\,M_\odot$.

\section{Size-dependent sputtering}
\label{app:size_eff}

\begin{table*}[t!]
\caption{Grain density and critical parameters for shattering and vapourisation for all the grain materials considered in the model.\vspace{.2cm}}
\label{table:shat_vap_par}      
\centering          
\begin{tabular}{l | c c c c c c }
\hline\hline
Parameter & Al$_2$O$_3$ & Fe$_3$O$_4$ & MgSiO$_3$ & Mg$_2$SiO$_4$ & AC & SiO$_2$\\
\hline
$\rho_0$\,[g\,cm$^{-3}$] & 3.83$^{\rm a}$ & 5.12$^{\rm a}$ & 3.18$^{\rm b}$ & 3.32$^{\rm b}$ & 2.2 & 2.65$^{\rm c}$ \\
$c_0$\,[km\,s$^{-2}$] & 6.91$^{\rm a}$ & 6.7$^{\rm a}$ & 6.0$^{\rm b}$ & 6.0$^{\rm b}$ & 1.8 & 3.68$^{\rm 3}$  \\
$s$ & 1.44$^{\rm a}$ & 1.36$^{\rm a}$ & 1.13$^{\rm b}$ & 0.86$^{\rm b}$ & 1.23  &2.12$^{\rm c}$ \\
\hline
\multicolumn{7}{c}{Shattering}  \\
\hline
$\epsilon_{\rm th}^{\rm sh}$\,[$\times 10^{10}$\,erg\,g$^{-1}$] & 32.4 & 6.13 & 105 & 105 & 0.08 & 6.31 \\
$P_{\rm th}^{\rm sh}$\,[$\times 10^{11}$\,dyn\,cm$^{-2}$] & 37.0$^{\rm d}$ & 20.8$^{\rm e}$ & 9.0$^{\rm f}$ & 9.0$^{\rm f}$ & 0.2 & 6.0$^{\rm g}$ \\
$v_{\rm th}^{\rm sh}$\,[km\,s$^{-1}$] & 16.1 & 7.0 & 28.9 & 28.9 & 0.9 & 7.1 \\
\hline
\multicolumn{7}{c}{Vapourisation}  \\
\hline
$\epsilon_{\rm th}^{\rm v}$\,[$\times 10^{10}$\,erg\,g$^{-1}$] & 8.1$^{\rm h}$ & 9.6$^{\rm i}$ & 7.7$^{\rm j}$ & 7.7$^{\rm j}$ & 64.0 & 30.0$^{\rm i}$ \\
$P_{\rm th}^{\rm v}$\,[$\times 10^{11}$\,dyn\,cm$^{-2}$] & 1.7 & 6.8 & 5.4 & 5.4 & 58.0 & 23.0 \\
$v_{\rm th}^{\rm v}$\,[km\,s$^{-1}$] & 8.1 & 8.8 & 7.8 & 7.8 & 23.0 & 15.0 \\
\hline
\end{tabular}
\\\vspace{2mm} 
References: $^{\rm a}$ \cite{LASL}, 
$^{\rm b}$ \cite{JGR:JGR5665}, $^{\rm c}$ \cite{2007M&PS...42.2079M}, $^{\rm d}$ \cite{Al2O3}, $^{\rm e}$ \cite{oxide}, $^{\rm f}$ \cite{Geodynamics}, $^{\rm g}$ IEEE Transactions on electron devices, 
$^{\rm h}$ \cite{scanning}, $^{\rm i}$ JANAF tables, $^{\rm j}$ \cite{2000MNRAS.318..809M}.
\end{table*}

With the use of the {\it Stopping and Range of Ions in Matter} 
software\footnote{SRIM is a collection of software packages which calculates the
transport of ions in matter, calibrating the results on the available
experimental data.} (SRIM, \citealt{ziegler}) it is possible to simulate the sputtering
of targets of different materials as they are hit by projectile ions.
The simulations provide both the sputtering yield (for a semi-infinite target), $Y_{\rm \infty}$, and 
the position ($x, y$ and $z$ coordinates) of the impinging ions after their interaction with the target.

The target materials and densities used in GRASH\_rev are listed in Table~\ref{table:shat_vap_par}. 
We consider He ions with an energy of 1 keV as projectiles.
In order to estimate the position of the displaced target atoms we follow the same approximation adopted by
\cite{2008A&A...492..127S}: a projectile loses most of its energy at a depth of $R_{\rm D} \sim 0.7 R_{\rm p}$.
We then assume that target atoms are displaced to a distance of $R_{\rm D}$ from their original position. In this
way, we can approximately reconstruct the distribution of displaced target atoms.

\begin{figure}[h!]
\begin{center}
\includegraphics[width=0.5\textwidth]{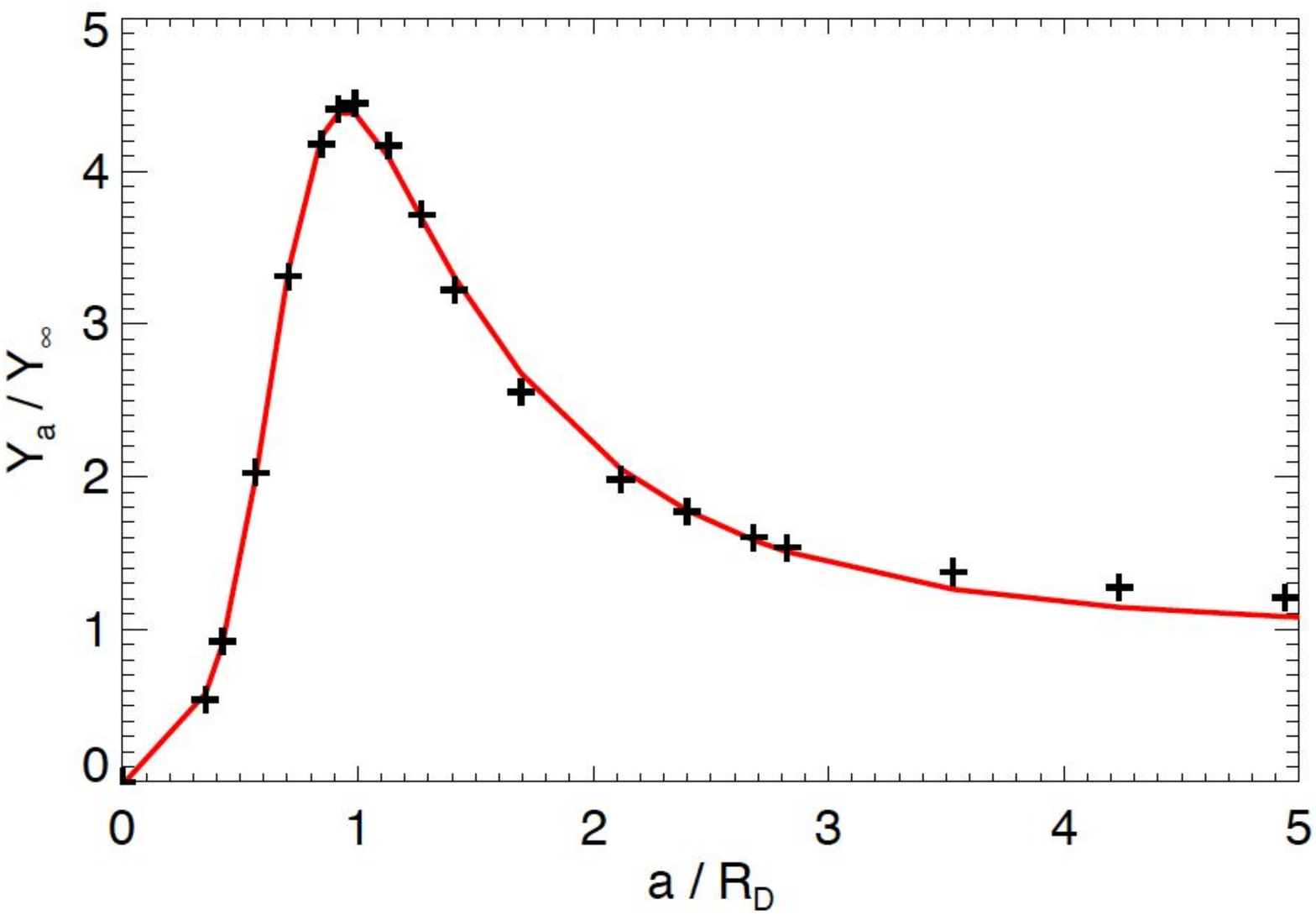}
\caption{Ratio $Y_{\rm a} / Y_{\infty}$ as a function of $x = a / R_{\rm D}$ for graphitic-type target material and for impinging He ions.
Plus signs represent results from SRIM simulations while the red solid line indicates the analytical approximation.}
\label{fig:size_eff}
\end{center}
\end{figure}

We simulate the bombardment of $n_{\rm TOT} = 10^6$ ions for each material.
At the end of the simulation we then have the 3D coordinates of the impinging ions and targets.
We consider different grain radii, $a$, and we count the number of impinging ions, $n_{\rm i}$, which are
enclosed in a sphere of radius $2a$ around the grain centre.
In this way we only account for  ions that triggered a collisional cascade in the target and exclude transmitted ions.
Furthermore, we count the number of targets, $n_{\rm t}$, whose position is outside the grain volume and
that are associated to ions that triggered collisional cascades in the target.
We then estimate the correction to the sputtering yield due to the effect of the grain size as:
\begin{equation}
Y_{\rm a} = \left(\frac{n_{\rm i} - n_{\rm t}}{n_{\rm TOT}} Y_{\infty} + \frac{n_{\rm t}}{n_{\rm TOT}} \right).
\end{equation}
In Fig.~\ref{fig:size_eff} we show the resulting ratio $f(x) = Y_{\rm a} / Y_{\infty}$ as a function of $x = a / R_{\rm D}$ (plus signs) 
for graphitic-type grains. 
We then fit our results (red line) with the parametric function in Eq.~\ref{eq:size_eff}.
The parameters resulting from the fit are reported in Table~\ref{table:size_eff} for all the materials.

\section{Shattering and vapourisation parameters}
\label{app:shat_vap}

Shattering and vapourisation parameters are calculated following the method by 
\cite{1994ApJ...431..321T} and \cite{1996ApJ...469..740J} and based upon material properties 
found in the literature.
The critical energy density for vapourisation, $\epsilon_{\rm th, v}$, is calculated starting from the surface binding energy, $E_{\rm b}$, 
as follows:
\begin{equation}
\epsilon_{\rm th, v} = 2 \frac{E_{\rm b}}{m_{\rm p} M},
\end{equation}
where $m_{\rm p}$ is the proton mass and $M$ is the molar mass of the considered material.
Then, the critical relative velocity between grains to set the vapourisation process is given by:
\begin{equation}
v_{\rm th, v} = 2 (2\epsilon_{\rm th}^{\rm v})^{2},
\end{equation}
and the corresponding pressure in the shocked solid is obtained as (\citealt{1994ApJ...431..321T}):
\begin{equation}
P_{\rm th, v} = \frac{\rho_0 v_{\rm th, v} (v_{\rm th, v} - c_0)}{s}.
\end{equation}
\noindent
The critical pressure, $P_{\rm th, sh}$, above which grains undergo fragmentation is assumed equal to 
the Young's modulus of the material.
The critical velocity for the onset of crater formation is given by (\citealt{1994ApJ...433..797J}):
\begin{equation}
v_{\rm th, sh} = \sqrt{4/3} \frac{P_{\rm th, sh}}{\rho_0 c_0},
\end{equation}
and the critical energy density results:
\begin{equation}
\epsilon_{\rm th, sh} = \frac{1}{2} \left(\frac{v_{\rm th,sh}}{2}\right)^2.
\end{equation}
In Table~\ref{table:shat_vap_par} we report all the values for 
vapourisation and shattering critical parameters that we have
adopted in GRASH\_rev.

\end{document}